\documentclass{aa}
\usepackage[dvips]{graphicx}
\usepackage{txfonts}
\usepackage{natbib}
\usepackage{color}
\bibpunct{(}{)}{;}{a}{}{,}    
\newcommand{\beq}{\begin{equation}}
\newcommand{\eeq}{\end{equation}}
\newcommand{\bea}{\begin{eqnarray}}
\newcommand{\eea}{\end{eqnarray}}
\newcommand{\lsim}{\raisebox{-0.6ex}{$\stackrel{{\displaystyle<}}{\sim}$}}

\newcommand{\url}[1]{{\tt #1}}

\def\gapp{\lower 3pt\hbox{${\buildrel > \over \sim}$}\ }
\def\lapp{\lower 3pt\hbox{${\buildrel < \over \sim}$}\ }

\newlength{\linwx}
\begin{document}
\title{Evolution of inclined planets in three-dimensional radiative discs}
\author{
Bertram Bitsch \inst{1}
and
Wilhelm Kley  \inst{1}
}
\offprints{B. Bitsch,\\ \email{bertram.bitsch@uni-tuebingen.de}}
\institute{
     Institut f\"ur Astronomie \& Astrophysik, 
     Universit\"at T\"ubingen,
     Auf der Morgenstelle 10, D-72076 T\"ubingen, Germany
}
\abstract
{While planets in the solar system only have a low inclination with respect to the ecliptic there is mounting
evidence that in extrasolar systems the inclination can be very high, at least for close-in planets.
One process to alter the inclination of a planet is through planet-disc interactions.
Recent simulations considering radiative transport have shown that the
evolution of migration and eccentricity can strongly depend on the thermodynamic state of the disc.
So far, this process has only been studied for a few selected planet masses using isothermal discs.}
{We extend previous studies to investigate the planet-disc interactions of fixed and moving planets
on inclined and eccentric orbits. We also analyse the effect of the disc's thermodynamic properties
on the orbital evolution of embedded planets in detail.
}
{The protoplanetary disc is modelled as a viscous gas where the internally produced dissipation
is transported by radiation. To solve the equations we use an explicit three-dimensional (3D) hydrodynamical code
{{\tt NIRVANA}} that includes full tensor viscosity, as well as implicit radiation transport in the
flux-limited diffusion approximation. To speed up the simulations we apply the {\tt FARGO}-algorithm in a 3D context.
}
{
For locally isothermal discs, we confirm previous results and find inclination damping and inward migration for planetary cores. For low inclinations ($i \lsim 2 H/r$), the damping is exponential, while it follows $di/dt \propto i^{-2}$ for larger $i$. For radiative discs, the planetary migration is very limited, as long as their inclination exceeds a certain threshold. If the inclination is damped below this threshold, planetary cores with a mass up to $\approx 33 M_{Earth}$ start to migrate outwards, while larger cores migrate inwards right from the start. The inclination is damped for all analysed planet masses.
}
{
In a viscous disc an initial inclination of embedded planets will be damped for all planet masses. This
damping occurs on timescales that are shorter than the migration time.
If the inclination lies beneath a certain threshold, the outward migration in radiative discs is not handicapped. 
However, only planets with a mass up to $\approx 33 M_{Earth}$ are prone to this outward migration.
Outward migration is strongest for circular and non-inclined orbits.
}
\maketitle
\markboth
{Bitsch \& Kley: Evolution of inclined planets in three-dimensional radiative discs}
{Bitsch \& Kley: Evolution of inclined planets in three-dimensional radiative discs}

\section{Introduction}
\label{sec:introduction}
Planets in our solar system do not all move in the same plane, but are inclined with respect to the ecliptic. 
The inclinations are in general small. The highest inclined planet is Mercury ($i=7.01^\circ$), and only some of the dwarf planets are inclined much higher (e.g. Pluto at $i=17^\circ$, or Eris at $i=44.2^\circ$). This low
inclination with respect to the ecliptic is typically taken as an indication that planets form within a flattened protoplanetary disc. The observed high $i$ for smaller objects are then a result of gravitational scattering processes
in the early solar system evolution.
Recently, it has been discovered that a relatively large number of transiting close-in planets in 
extrasolar systems can be inclined significantly with respect to the stellar rotation axis
\citep[see e.g.][]{2010arXiv1008.2353T}.
This discovery makes it necessary to take inclination into consideration in both numerical simulations and 
theoretical considerations of planet-disc interaction.

Starting from early work \citep{1980ApJ...241..425G, 1984ApJ...285..818P, 1986Icar...67..164W},
the effect a protoplanetary disc has on the migration of the planets has been analysed
intensively after the discovery of hot Jupiters close to the star (for a review see
\citet{2007prpl.conf..655P}). Numerical simulations of planet-disc interactions are
a valuable tool for a better understanding. Two-dimensional (2D) studies have been performed for example by \citet{1999ApJ...514..344B, 2002A&A...385..647D} and later fully three dimensional (3D) simulations \citep{2003ApJ...586..540D, 2003MNRAS.341..213B} followed.
All these simulations only considered isothermal discs. However, recent studies (starting with the work of \citet{2006A&A...459L..17P}) have shown that including radiation transport in planet-disc interaction studies resulted in a slowed down or even reversed migration \citep{2008ApJ...672.1054B, 2008A&A...485..877P, 2008A&A...478..245P, 2008A&A...487L...9K, 2010MNRAS.408..876A}. All authors agree that the inclusion of radiative transfer is important and that it strongly affects the migration properties of low mass planets (type-I-migration) in discs, even though there are differences in the magnitude of the effect. It also affects whether it actually leads to reversal of migration or only to a slow down.

Recently, we have analysed the disc's influence on the orbital eccentricity \citep{2010A&A.523...A30} for planets of
different masses in fully radiative discs. We showed that planet-disc interaction leads quite generally 
(for all planet masses) to a damping
of planetary eccentricity on shorter timescales than the migration timescale.
Additionally, planets on eccentric orbits embedded in fully radiative discs only migrate outwards, when the eccentricity is damped to a nearly circular orbit. Clearly, eccentricity destroys the very sensitive corotation torques near the planet, which results in inward migration until the orbit is circularised again.

The effect of disc-planet interaction on inclination has not yet been studied in any wide extent, owing to the heavy computational requirements.  \citet{2004ApJ...602..388T} analysed the influence of the disc on the inclination of low mass planets in linear studies. They find an exponential damping for any non-vanishing inclination, but their results are only formally valid for $i << H/r$. Numerical simulations of inclined planets with higher inclinations than $H/r$ (inclination in radians) have shown that exponential damping may be valid up to $i \approx 2 H/r$. For planets on higher initial inclined orbits, the damping rate deviates from being exponential, and it can be fitted best with a $di/dt \propto i^{-2}$ function \citep{2007A&A...473..329C}.
Only low mass planets with a mass of about 20-30 earth masses were considered in these studies. The influence on an
inclined Jupiter type planet has been considered by \citet{0004-637X-705-2-1575}.
They find that highly inclined and eccentric planets with Jovian masses lose their inclination and eccentricity very fast (on a timescale of order of $10^3$ years), when entering the disc again.
Since a highly inclined planet is only disturbed in a small way by the accretion disc (and vice versa), such a
planet is only able to open a gap in the disc, when the inclination drops to $i<10.0^\circ$. 
\citet{2010arXiv1001.0657T} considered inclination evolution utilising a Kozai-type of effect for planets on high inclined and eccentric orbits in the presence of the disc.

In the present work we will extend the work on inclined planets and analyse the effect of planet-disc interaction on embedded planets of all masses. In fully three-dimensional hydrodynamical simulations we will consider isothermal as well as fully radiative discs and study the change in inclination of an embedded planet.

In Section \ref{sec:model} we give a short overview of our code and numerical methods. In Section \ref{sec:fixedplanets} we measure the change of inclination and semi-major axis for $20 M_{Earth}$ planets on fixed circular inclined orbits and then let these planets evolve their orbits in the disc due to the torque acting on the planet form the disc in Section \ref{sec:moveplanets}. In Section \ref{sec:massplanets} we study the influence of the planetary mass on the evolution of planets on inclined orbits. In Section \ref{sec:inceccplanets} we follow the evolution of $20 M_{Earth}$ planets on eccentric and inclined orbits. In Section \ref{sec:fixedplanets} to Section \ref{sec:inceccplanets} we also point out the differences between the isothermal and fully radiative simulations. Finally we summarise and conclude in Section \ref{sec:Sumcon}.

\section{Physical modelling}
\label{sec:model}

The protoplanetary disc is modelled as a three-dimensional (3D), non-self-gravitating gas whose motion is described by the Navier-Stokes equations. We treat the disc as a viscous medium, where the dissipative effects can then be described via the standard viscous stress-tensor approach \citep[eg.][]{1984frh..book.....M}. We also assume that the heating of the disc occurs solely through internal viscous dissipation and ignore the influence of additional energy sources (e.g. irradiation form the central star). This internally produced energy is then radiatively diffused through the disc and eventually emitted from its surface. For this process we use the flux-limited diffusion approximation \citep[FLD,][]{1981ApJ...248..321L}, which allows us to treat the transition from optically thick to thin regions as an approximation. A more detailed description of the modelling and the numerical methodology is provided in our previous papers \citep{2009A&A...506..971K} and \citep{2010A&A.523...A30}.
Compared to our previous papers, we now extend the simulations by including planets with different masses on inclined orbits.

\subsection{General Setup}
An important issue in modelling planetary dynamics in discs is the gravitational potential of the planet since this has to be artificially smoothed to avoid singularities. 

While in two dimensions a potential smoothing takes care of the otherwise neglected
vertical extension of the disc, in three dimensional simulations the most accurate
potential should be used. As the planetary radius is much smaller than a typical grid cell, and the planet is treated as a point mass, a smoothing of the potential is required to ensure numerical stability.
In  \citet{2009A&A...506..971K}  we have discussed two alternative prescriptions for
the planetary potential. The first is the classic $\epsilon$-potential 
\begin{equation}
\label{eq:epsilon}
   \Phi_p^{\epsilon} = - \frac{G m_p}{\sqrt{d^2 + \epsilon^2}} \, .
\end{equation}
Here $m_P$ is the planetary mass, and $d=| \mathbf{r} - \mathbf{r_P}|$ denotes the distance of the disc element to the planet.
This potential has the advantage that it leads to very stable evolutions when the parameter $\epsilon$ (stated in units of the Hill radius) is not
too small, i.e. is a significant fraction of the Roche-lobe size. 
The disadvantage is that for smaller $\epsilon$, which would yield a higher accuracy at larger $d$, the potential becomes very deep at the planetary position.
Additionally, due to the long range nature, the potential differs from the exact one even for medium to larger distances $d$
from the planet. 

To resolve these problems at small and large $d$ simultaneously, we have suggested 
the following {\it cubic}-potential \citep{2006A&A...445..747K,2009A&A...506..971K} 
\begin{equation}
\label{eq:cubic}
\Phi_p^{cub} =  \left\{
    \begin{array}{cc} 
   - \frac{m_p G}{d} \,  \left[ \left(\frac{d}{r_\mathrm{sm}}\right)^4
     - 2 \left(\frac{d}{r_\mathrm{sm}}\right)^3 
     + 2 \frac{d}{r_\mathrm{sm}}  \right]
     \quad &  \mbox{for} \quad  d \leq r_\mathrm{sm}  \\
   -  \frac{m_p G}{d}  \quad & \mbox{for} \quad  d > r_\mathrm{sm} 
    \end{array}
    \right.
\end{equation}
Here, $r_{sm}$ is the smoothing length of the potential in units of the Hill radius. The construction of the planetary potential is in such
a way that for distances larger than $r_{sm}$ the potential matches the correct $1/r$ potential.
Inside this radius ($d < r_{sm}$) it is smoothed by a cubic polynomial. This potential has the advantage of exactness
outside the specified distance $r_{sm}$, while being finite inside. The parameter $r_{sm}$ is equal to $0.5$ in all our simulations, unless stated otherwise. 
The value of the smoothing length of the planetary potential $r_{sm}$ was discussed in great detail in \citet{2009A&A...506..971K}, 
where we compared different smoothing lengths and potential types, resulting in our choice of $r_{sm}=0.5$. 
In this work we use the cubic-form for the planetary potential for all fully radiative simulations and for the isothermal simulations with planets on fixed orbits.
We note that for $\epsilon=0.25$ the depth of the $\epsilon$-potential at the position of the planet, $d=0$, is identical to the
cubic-potential with $r_{sm} =0.5$. However, the difference reaches a maximum of about 20\% at $d=0.25$, is 10\% at $d=0.5$, and slowly diminishes for larger $d$.

Our simulations for moving planets in isothermal discs have shown that in this case the $r_{sm} =0.5$ cubic potential yields unstable
evolutions. This is due to the relatively steep centre which leads to a significant density enhancement for isothermal discs. 
Hence, we used in this case directly the much smoother $\epsilon$-potential with $\epsilon=0.8$ instead, which has been used before in a similar
context \citep{2007A&A...473..329C}, without having explored other choices here.

The gravitational torques and forces acting on the planet are calculated by integrating over the whole disc, where we apply a tapering function to exclude the inner parts of the Hill sphere of the planet. Specifically, we use the smooth (Fermi-type) function
\begin{equation}
\label{eq:fermi}
     f_b (d)=\left[\exp\left(-\frac{d/R_H-b}{b/10}\right)+1\right]^{-1}
\end{equation}
which increases from 0 at the planet location ($d=0$) to 1 outside $d \geq R_{H}$ with a midpoint $f_b = 1/2$ at $d = b R_{H}$, i.e. the quantity $b$ denotes the torque-cutoff radius in units of the Hill radius. This torque-cutoff is necessary to avoid large, probably noisy contributions from the inner parts of the Roche lobe and to disregard material that 
is possibly gravitationally bound to the planet \citep{2009A&A...502..679C}.
Here we assume (as in our previous paper) a transition radius of $b= 0.8$ Hill radii,
as a change in $b$ did not influence the results significantly \citep{2009A&A...506..971K}.

\subsection{Initial Setup}

The three-dimensional ($r, \theta, \phi$) computational domain consists of a complete annulus of the protoplanetary disc centred on the star, extending from $r_{min}=0.4$ to $r_{max}=2.5$ in units of $r_0=a_{Jup}=5.2 AU$. In the vertical direction the annulus extends $7^\circ$ below and above the disc's midplane, meaning $83^\circ < \theta < 97^\circ$. Here $\theta$ denotes the polar angle of our spherical polar coordinate system measured from the polar axis.
The central star has one solar mass $M_\ast = M_\odot$, and the total disc mass inside [$r_{min}, r_{max}$] is $M_{disc} = 0.01 M_\odot$. For the isothermal simulations we assume an aspect ratio of $H/r=0.037$ for the disc, in very close agreement with the
fully radiative models of our previous studies. For the radiative models $H/r$ is calculated self-consistently from the equilibrium structure
given by the viscous internal heating and radiative diffusion.
The isothermal models are initialised with constant temperatures on cylinders with a profile $T(s) \propto s^{-1}$ with $s=r \sin \theta$. This yields a constant ratio of the disc's vertical height $H$ to the radius $s$. The initial vertical density stratification is approximately given by a Gaussian:
\begin{equation}
  \rho(r,\theta)= \rho_0 (r) \, \exp \left[ - \frac{(\pi/2 - \theta)^2 \, r^2}{2 H^2} \right] \ .
\end{equation} 
Here, the density in the midplane is $\rho_0 (r) \propto r^{-1.5}$ which leads to a $\Sigma(r) \propto \, r^{-1/2}$ profile of the vertically integrated surface density. In the radial and $\theta$-direction we set the initial velocities to zero, while for the azimuthal component the initial velocity $u_\phi$ is given by the equilibrium of gravity, centrifugal acceleration and the radial pressure gradient. This corresponds to the equilibrium configuration for a purely isothermal disc. 

For our fully radiative model we first run a 2D axisymmetric model (starting from the given isothermal equilibrium) to obtain a new self-consistent equilibrium where viscous heating balances radiative transport/cooling from the surfaces. After reaching that equilibrium, we extend this model to a full 3D simulation, by expanding the grid into $\phi$-direction. The resulting disc for this model has $H/r \approx 0.037$ so we choose that value
for our isothermal runs.

\subsection{Numerical Setup}

Our coordinate system rotates at the initial orbital frequency of the planet (at $r=r_0$). We use an equidistant grid in $r,\theta,\phi$ with a resolution of ($N_r,N_\theta,N_\phi)=(266,64,768)$ active cells for our simulations. At $r_{min}$ and $r_{max}$ we use damping boundary conditions for all three velocity components to minimise disturbances (wave reflections) at these boundaries. The velocities are relaxed towards their initial state on a timescale of approximately the local orbital period. The angular velocity is relaxed towards the Keplerian values, while the radial velocities at the inner and outer boundaries vanish. Reflecting boundary conditions are applied for the density and temperature in the radial directions. We apply periodic boundary conditions for all variables in the azimuthal direction. In the vertical direction we set outflow boundary conditions for $\theta_{min} = 83^\circ$ and $\theta_{max} = 97^\circ$ (the surfaces of the disc). 
We use the finite volume code {\tt NIRVANA} \citep{1997ZiegYork} with implicit radiative transport in the flux-limited
diffusion approximation and the {\tt FARGO} \citep{2000A&AS..141..165M} extension as described in \citet{2009A&A...506..971K}.

\subsection{Simulation Setup}

In the first part of our model sequence we consider the orbital evolution of a planet with a fixed mass ($20 M_{Earth}$) on inclined orbits using different initial inclinations. For comparison we consider isothermal and fully radiative models. Using radiative discs here is a direct extension of a previous study under purely isothermal disc conditions using the same planet mass \citep{2007A&A...473..329C}. We distinguish two different approaches for these $20 M_{Earth}$ models: first, a model sequence where the planet stays on a fixed inclined orbit and secondly where the planet is free to move inside the computational domain under the action of the planet-disc gravitational forces. For the second models we insert the planet in the disc and let it move immediately, but using a time-dependent mass growth of the planet (through the planetary potential) with a typical switch on time of $10$ planetary orbits. For the first set of models the $20 M_{Earth}$ planet is inserted at the
nominal mass in the disc at the start of the simulation. Initially the planet starts at a distance $r=a_{Jup}=5.2 AU$ from the central star.  For the fully radiative simulations we set the ambient temperature to a fixed value of $10 K$ at the disc surface (at $\theta_{min}$ and $\theta_{max}$), which ensures that all the internally generated energy is liberated freely at the disc's surface. This low temperature boundary condition works very well at optically thin boundaries and does not effect the inner parts of the optically thick disc \citep{1999ApJ...518..833K, 2009A&A...506..971K}. In the second part of the project we consider sequences of models for a variety of planet masses. We note that a $20 M_{Earth}$ planet has in our simulations, using our standard resolution, a Roche radius of about $3.3$ grid cells. In the last part we limit ourselves to a $20 M_{Earth}$ planet and simulate the evolution of planets with an initial eccentricity and inclination. For all isothermal simulations, which allow the planet to move freely inside the disc, we use the $\epsilon$-potential. Only for isothermal simulations of planets on fixed inclined orbits, the cubic potential is used. Additionally, the cubic potential is used for all fully radiative simulations as well.

\section{Models with an embedded planet on fixed circular inclined orbits}
\label{sec:fixedplanets}

In this section we consider planets on fixed circular and inclined orbits embedded in either isothermal or fully radiative discs. From the disc forces acting on the planet on the fixed orbit we calculate its change of inclination and its migration rate. In this section all simulations use the cubic potential, featuring $M_{Planet} = 20 M_{Earth}$, and a semi-major axis of $a=1.0$. Our simulations only cover up to $7^\circ$ above and below the equatorial plane, as the disc gets very thin for these regions. However, we investigate the motion of planets for higher inclinations (for fixed planets up to $15^\circ$). The results are not influenced by our limited vertical extent of our computational grid, as the density is very low in the upper layers of the disc.

\subsection{Change of inclination}

To determine the change of orbital elements for planets on fixed inclined orbits, we follow \citet{burns:944}. If a small disturbing force $\mathbf{dF}$ (given per unit mass) due to the disc is acting on the planet, the planet changes its orbit. This small disturbing force $\mathbf{dF}$ may change the planetary orbit in size (semi-major axis $a$), eccentricity $e$ and inclination $i$. The inclination $i$ gives the angle between the orbital plane and some arbitrary fixed plane, which is in our case the equatorial ($\theta = 90^\circ$) plane, which corresponds to the midplane of the disc. Only forces lying in the orbit plane can change the orbits size and shape, while these forces can not change the orientation of the orbital plane. In \citet{burns:944} the specific disturbing force is written as
\begin{equation}
\label{eq:distforce}
     \mathbf{dF} = \mathbf{R} + \mathbf{T} + \mathbf{N} = R \mathbf{e}_{R} + T \mathbf{e}_T + N \mathbf{e}_N \ ,
\end{equation}
where the $\mathbf{e}$'s represent an orthogonal unit vector triad. The perturbing force can be split in these components: $\mathbf{R}$ is radially outwards along $\mathbf{r}$, $\mathbf{T}$ is transverse to the radial vector in the orbit plane (positive in the direction of motion of the planet), and $\mathbf{N}$ is normal to the orbit planet in the direction $\mathbf{R} \times \mathbf{T}$.

\citet{burns:944} finds for the change of inclination
\begin{equation}
\label{eq:ichange}
     \frac{di}{dt} = \frac{a N \cos \xi}{H} \ ,
\end{equation}
where the numerator is the component of the torque which rotates the specific angular momentum $\mathbf{H} = \mathbf{r} \times \mathbf{\dot{r}}$ about the line of nodes (and which thereby moves the orbit plane). The specific angular momentum $H$ is defined as
\begin{equation}
	H = \sqrt{ GM a (1 - e^2)} \ .
\end{equation}

The angle $\xi$ is related to the true anomaly $f$ in the following way $f = \xi - \omega$, with $\omega$ being the argument of periapsis. $\xi$ describes the angle between the line of nodes and the planet on its orbit around the star. For our case of circular orbits the argument of periapsis $\omega$ is zero.

\begin{figure}
 \centering
 \includegraphics[width=0.9\linwx]{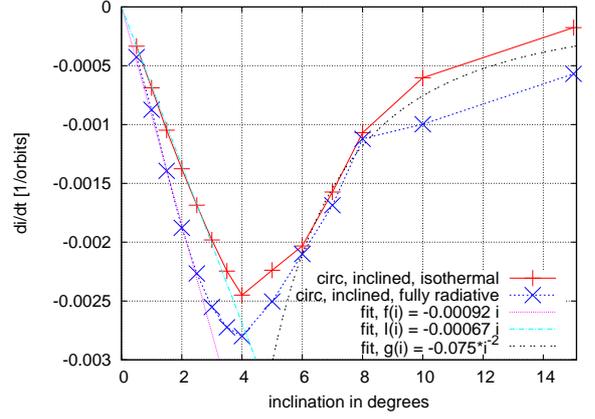}
 \caption{Calculated rate of change of the inclination ($\frac{di}{dt}$) for inclined $20 M_{Earth}$ planets on fixed circular orbits. Results for the isothermal (red) and fully radiative simulations (dark blue) are shown. Overlayed are in light blue and purple fits for the exponential decay of inclination for small inclinations and in black for the decay $di/dt \propto i^{-2}$ for larger inclinations.
   \label{fig:Incchangeplot}
   }
\end{figure}

In the simulations the planet's inclination spans from $i=0.5^\circ$ to $i=15.0^\circ$, in the isothermal as well as in the fully radiative regime. In Fig.~\ref{fig:Incchangeplot} we display the rate of change of inclination for planets on fixed inclined orbits in the isothermal and fully radiative scheme, after we average $di/dt$ over $2$ planetary orbits, after $150$ orbits. As the angle $\xi$ changes after every time step, one needs to average the quantity of $di/dt$ over the time of a planetary orbit to determine an exact value for the change of inclination of the planetary orbit. For both thermodynamic systems the change in inclination is nearly identical and always negative, meaning that moving planets with a non zero inclination lose their initial inclination at a rate according to the current inclination. At $i=4.0^\circ$ the loss of inclination is at it's maximum and is reduced for higher and lower inclinations. For high inclined planets ($i > 6.0^\circ$) this loss is $di/dt \propto i^{-2}$, as indicated by the short-dashed fit in black in Fig.~\ref{fig:Incchangeplot}, and for lower inclined planets ($i < 4.0^\circ$) the inclination is damped with an exponential decay as indicated by the linear fits in light blue and purple in Fig.~\ref{fig:Incchangeplot}. 
This is identical to the behaviour found in \citet{2007A&A...473..329C} and is also confirmed by our simulations of moving planets below.

From our measured $di/dt$ for planets on fixed orbits, we can calculate the inclination damping timescale $\tau_{inc}$ according to
\begin{equation}
 \frac{\overline{di/dt}}{i}= - \frac{1}{\tau_{inc}}
\end{equation}
with $i$ being the inclination of the planet. In the exponential, small $i$ regime we obtain for the isothermal simulations $\tau_{inc}=27$ orbits and for the fully radiative simulations $\tau_{inc}=20$ orbits. Below we compare this to the linear results of \citet{2004ApJ...602..388T} and moving planets.

\subsection{Change of semi-major axis}

The power on the planet determines the change in semi-major axis of the planet, while the torque represents a change in both eccentricity and semi-major axis \citep{2010A&A.523...A30}. For circular orbits torque and power are identical. 
For that reason we display and refer only to the torque acting on the planet, instead of mentioning the power as well. 
A positive torque indicates outward migration, while a negative torque represents inward migration.

\begin{figure}
 \centering
 \includegraphics[width=0.9\linwx]{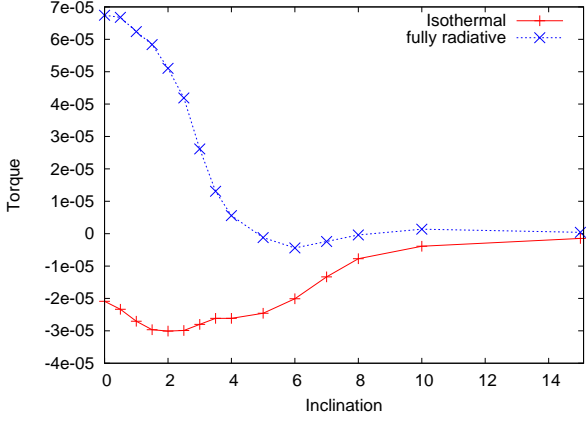}
 \caption{Torque acting on a $20 M_{Earth}$ planet on a fixed inclined circular orbit in dependency of the inclination and thermodynamics of the disc. Torque and power are the same for planets on circular orbits. The torque has been averaged over $20$ orbits, taken from $t=180$ to $t=200$ orbits for the isothermal (solid red line) and from $t=120$ to $t=140$ for the fully radiative simulation (dashed blue line).
 \label{fig:InctorIsofull}
 }
\end{figure}

In Fig.~\ref{fig:InctorIsofull} we display the torque acting on the planet for both thermodynamic systems. The torque is positive for the fully radiative disc and negative for the isothermal disc. In the fully radiative scheme the torque has its maximum at $i=0.0^\circ$ with a massive loss towards higher inclinations until about $i \approx 4.5^\circ$, when the torque is about zero. For even higher inclinations the torque remains at about zero level. In the isothermal case the torque is negative and nearly constant for all inclinations, but with an increase towards higher inclinations.

As for circular orbits the torque determines the change of semi-major axis for our embedded planets, we expect a positive migration rate for planets in a fully radiative disc and a negative migration rate for planets in an isothermal disc. To determine a migration rate for planets on fixed orbits, we follow \citet{burns:944} again:
\begin{equation}
	\frac{da}{dt} = \frac{2}{(G M_*)^{1/2}} a^{3/2} (1-e^2)^{-1/2} \left[ R \, e \sin f + T (1+e \sin f)\right] \ ,
\end{equation}
where $a$ is the semi-major axis, $e$ the eccentricity, $f = \xi - \omega$ the true anomaly and $R$ and $T$ are radial and tangential directions of the disturbing force, see eq.~(\ref{eq:distforce}). Please note that only forces lying in the orbit plane can change the orbit size, 
and that for circular orbits the specific total torque on the planet is simply given by $\Gamma_{tot} = a T$,
with $\dot{H} = \Gamma_{tot}$.

\begin{figure}
 \centering
 \includegraphics[width=0.9\linwx]{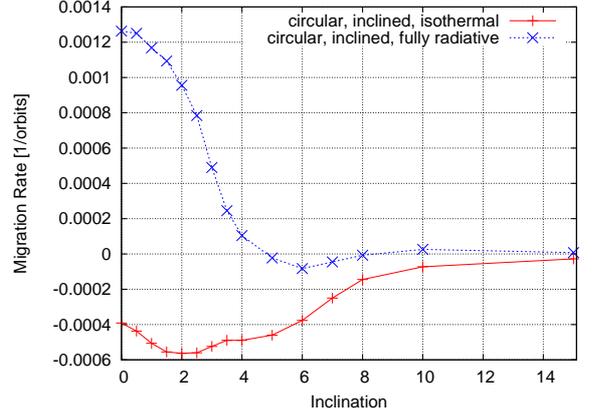}
 \caption{ Calculated rate of change of the semi-major axis ($\dot{a}/a$) for inclined $20 M_{Earth}$ planets on fixed, circular orbits. For inclinations lower than $i \leq 4.5^\circ$ we observe a positive migration rate, which suggest that low inclinations do not stop outward migration, while we observe a negative migration rate for the isothermal case.
 \label{fig:AIncIsofullrate}
 }
\end{figure}

The rate of migration is displayed in Fig.~\ref{fig:AIncIsofullrate}. In the fully radiative scheme the planet experiences a positive migration rate (outward migration) for all inclinations smaller than $i \leq 4.5^\circ$, but with a strong increase towards lower inclinations (about a factor of $10$ for the difference between $i=4.0^\circ$ and $i=1.5^\circ$). For inclinations lower than $i \approx 1.5^\circ$ the migration rate stays approximately constant. Planets with an inclination of $4.5^\circ$ and higher seem to be stalled and migrate neither inwards nor outwards. The effects responsible for outward migration seem to be only valid in the small inclined case. 

In the isothermal case the migration rate is nearly constant and negative with a slight increase towards higher inclinations, suggesting that an inclined planet will move inwards in the isothermal regime for low inclined planets. Higher inclined planets will move inwards only at a very small rate compared to their low inclined counterparts. For large $i$ the change of semi-major axis approaches zero, as the planet's orbit is high above the dense regions of the disc, which are capable of influencing the planet's orbit.

From these results we conclude that a planet with a non-zero initial inclination will lose this inclination in time. This inclination loss has in principal no effect on the trend of migration, so a planet in a fully radiative disc will migrate outwards, while a planet in an isothermal disc will migrate inwards. In one of the next chapters we will observe moving planets with a non-zero inclination, which will do exactly that.

\subsection{Torque analysis}

\begin{figure}
 \centering
 \includegraphics[width=0.9\linwx]{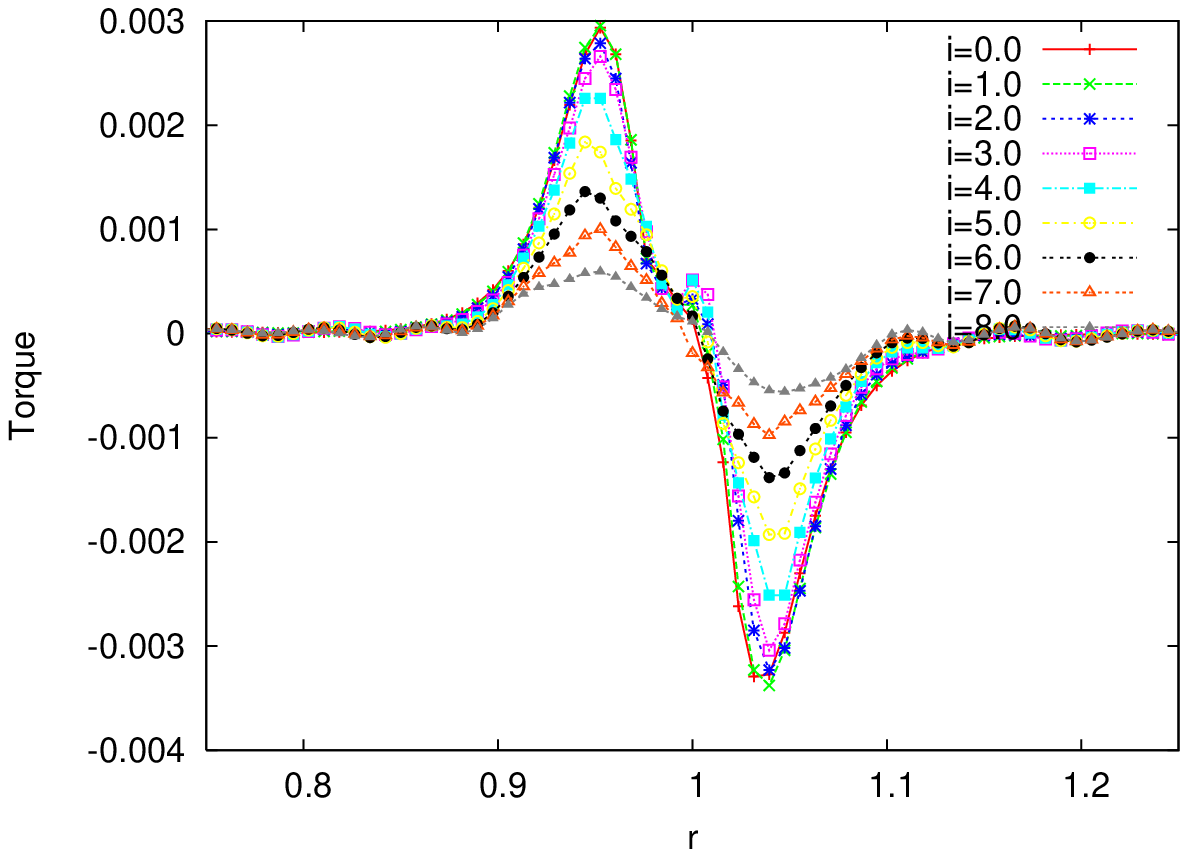}
 \includegraphics[width=0.9\linwx]{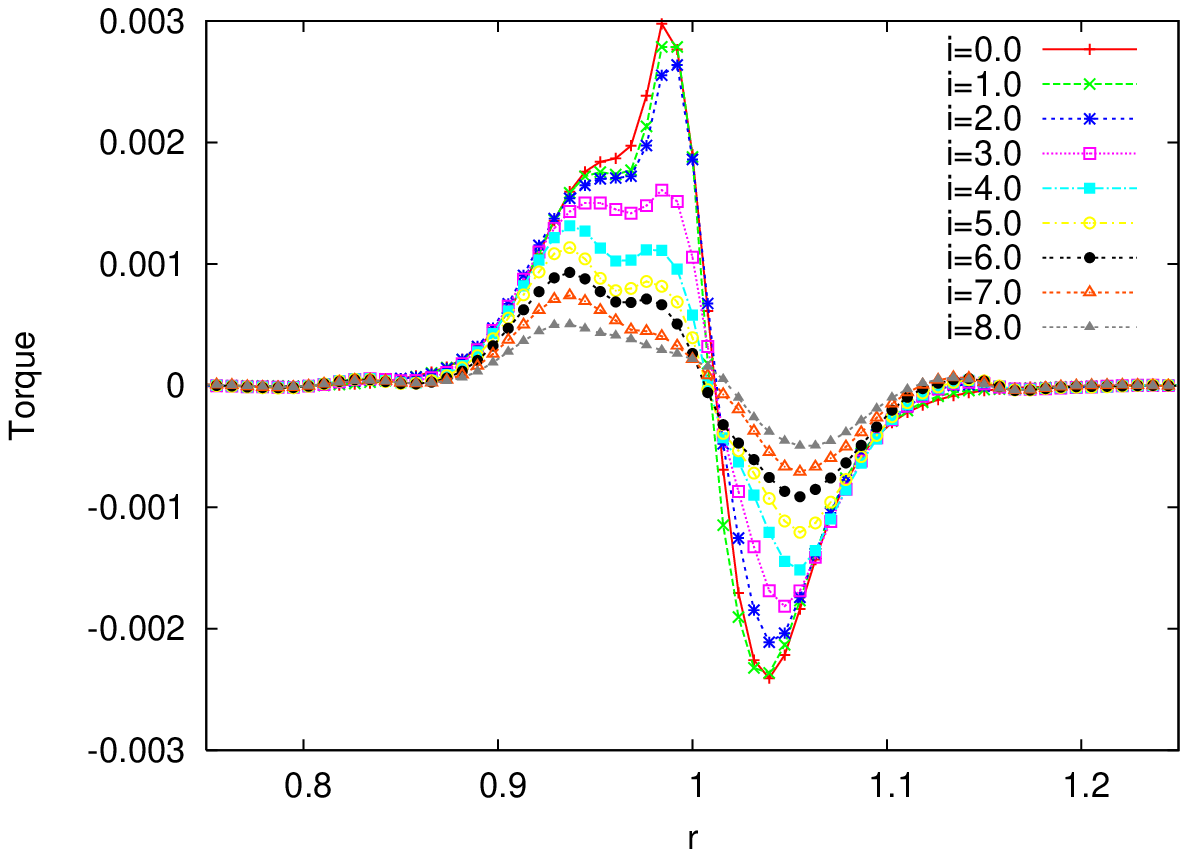} 
 \caption{Torque density acting on a planet on a fixed inclined orbit in dependency of the inclination of the planet embedded in an isothermal $H/r=0.037$ disc (top) and a fully radiative disc (bottom). The snapshots are taken at $t=90$ orbits for the isothermal simulations and at $t=150$ orbits for the fully radiative simulations. At this time planet-
disc interactions are in equilibrium and the planet is at its lowest point in orbit (lower culmination).
   \label{fig:IncGammaIsofull}
   }
\end{figure}

To understand the behaviour of the total torque in more detail we now analyse the space-time variation of the torque of the planet.
For that purpose we introduce the radial torque density $\Gamma (r)$, which is defined in such a way that the total torque $\Gamma_{tot}$ acting on the planet is given by
\beq
       \Gamma_{tot} = \int_{r_{min}}^{r_{max}} \, \Gamma(r) \, dr \ .
\eeq
The radial torque density has been a very useful tool to investigate the origin of the torques in our previous work on planets on fixed circular orbits. Even though the planet may be on an inclined orbit, this does not alter our definition of the radial torque density $\Gamma (r)$ which is calculated in any case with respect to the orbital plane of the planet.
In Fig.~\ref{fig:IncGammaIsofull} we display $\Gamma (r)$ for a selection of our planets in the isothermal and fully radiative regime.

\begin{figure}
 \centering
 \includegraphics[width=0.9\linwx]{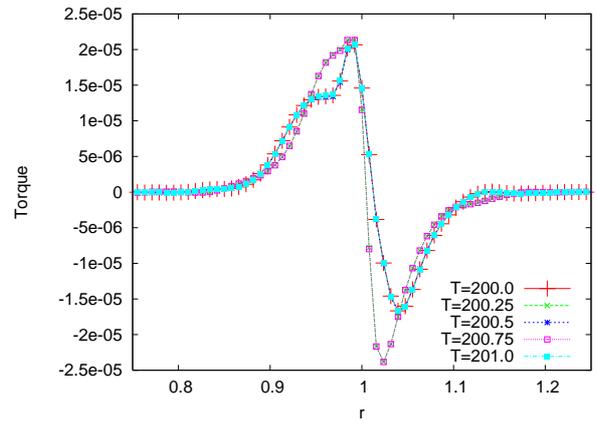}
 \caption{Torque density acting on a $20 M_{earth}$ planet on a fixed inclined orbit with $i=2.0^\circ$ in a fully radiative disc at different phases within its orbit. Integer (half integer) values of the indicated time refer to the upper (lower) culmination.
 At the intermediate quarterly orbit times the planet moves through the nodal line.
   \label{fig:Torquefull20}
   }
\end{figure}

\begin{figure}
 \centering
 \includegraphics[width=1.0\linwx]{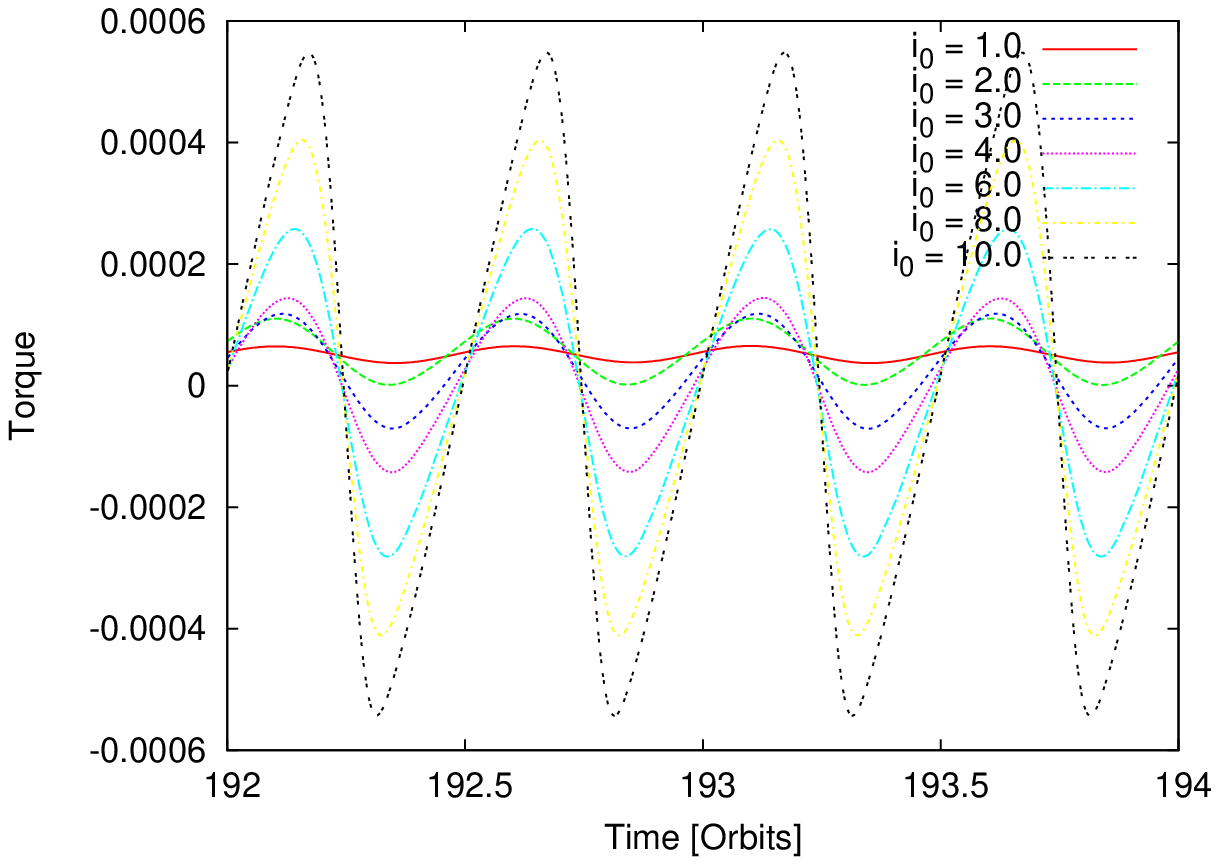}
 \includegraphics[width=1.0\linwx]{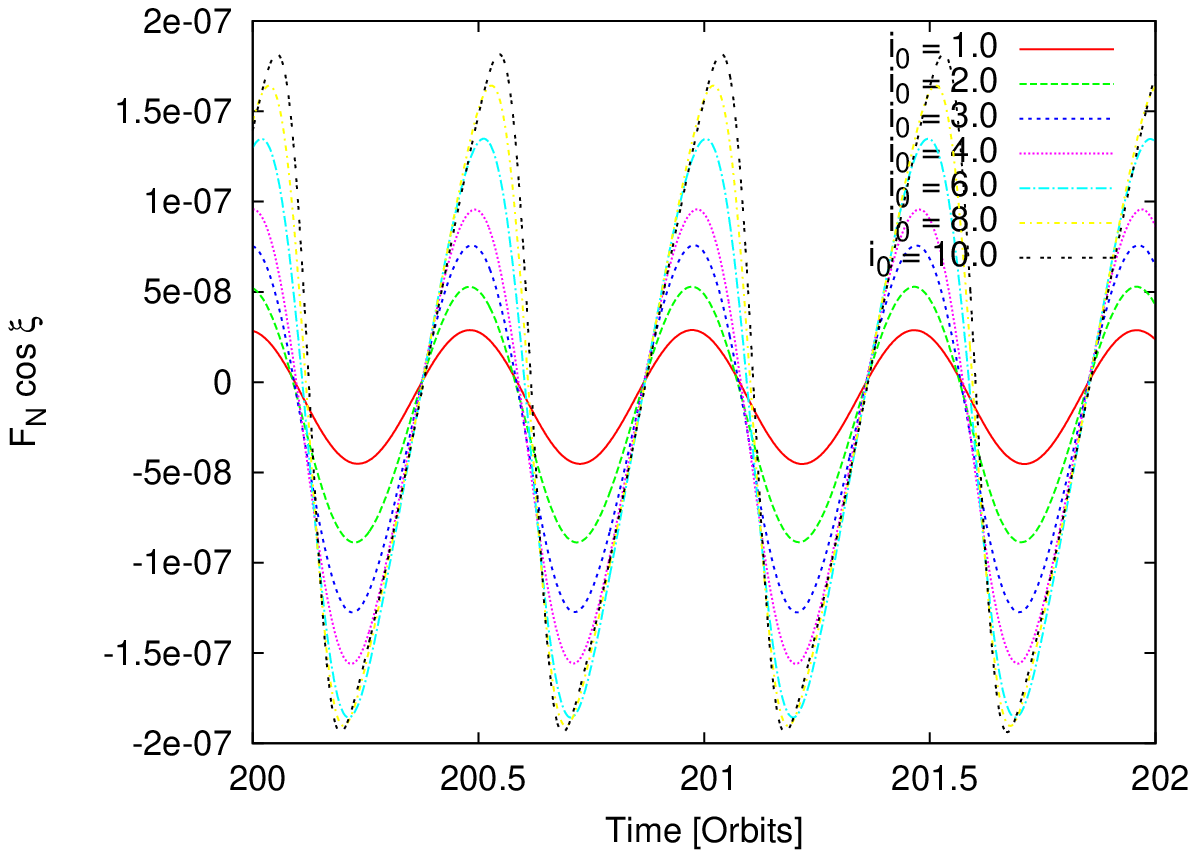}
 \caption{In the top panel we display the evolution of the torque acting on planets in a fully radiative disc on fixed circular, inclined orbits in time. In the bottom panel the normal component of the disturbing force $\mathbf{dF}$, ${N} \cos \xi$ as to be used eq.~(\ref{eq:ichange}), is displayed in dependency of time.
   \label{fig:TorqueFNtime}
   }
\end{figure}

In the isothermal case the $\Gamma (r)$-function follows the same trend for all displayed inclinations. The only difference is near the planet, around the $r=1.0$ region. The torque of the low inclined planets declines constantly, while the higher inclined planets ($2.0 \leq i \leq 4.0$) form a little spike in the $\Gamma (r)$ distribution, which disappears for even higher inclinations again. For higher inclinations the Lindblad-Torques become less and less pronounced, as can be seen in Fig.~\ref{fig:IncGammaIsofull} (top panel). 
Please keep in mind that for the inclined planets the torque density changes during one orbit, 
as can be seen from the change in the total torque acting on the planet (top panel in Fig.~\ref{fig:TorqueFNtime}). 
This is also illustrated in Fig.~\ref{fig:Torquefull20} for the planet with $i=2.0^\circ$. It can been seen that for
symmetry reasons $\Gamma(r)$ is identical at upper and lower culmination, and at the crossings of the nodal line,
respectively.

In Fig.~\ref{fig:TorqueFNtime} the torque acting on planets (please note that for circular orbits torque and power are identical) with different inclinations on fixed circular orbits (top) and the normal component of the disturbing force $\mathbf{dF}$ (bottom) for fully radiative discs is displayed. The torque acting on the planets oscillates with time, making two oscillations every orbit. The planet starts at the highest point in orbit (upper culmination) and evolves to the lowest point in orbit (lower culmination) in the time of half an orbit. Then the planet moves to the upper culmination again and reaches it at the end of the orbit. The planets with a lower inclination ($i< 4.0^\circ$) have clearly a positive average torque, while for the planets with higher inclinations positive and negative contributions approximately cancel out such that the average total torque is very small (see Fig.~\ref{fig:InctorIsofull}). Higher inclinations also trigger higher amplitudes in the torque distributions. These amplitudes in the torque distribution suggest oscillations in the evolution of the semi-major axis of the planet (to a small degree), which are visible when the planet is allowed to move freely inside the disc (see Fig.~\ref{fig:AIncIso}, Fig.~\ref{fig:AIncIsoHr005} and Fig.~\ref{fig:AIncfull}). 

In the lower panel in Fig.~\ref{fig:TorqueFNtime} the normal component of the disturbing force $N \cos \xi$ is displayed. $N \cos \xi$ oscillates also twice in every orbit, as the torque, but it is slightly shifted with respect to the torque. Negative values indicate a reduction of inclination, while a positive force indicates an increase of inclination. The oscillations in $N \cos \xi$ indicate oscillations in the inclination, when the planet is allowed to move freely in the disc. Indeed, in case the planet moves inside the disc, these oscillations in the inclination become visible at the beginning of the simulations (see Fig.~\ref{fig:IncIncIso}, Fig.~\ref{fig:AIncIsoHr005} and Fig.~\ref{fig:IncIncfull} below). On average the normal component of the disturbing force $N \cos \xi$ is negative, which indicates inclination damping for all simulated inclinations, see Fig.~\ref{fig:Incchangeplot}


For simulations in the isothermal regime, the results are nearly identical, with only one exception: The overall torque is negative, which results in inward migration. The normal component of the disturbing force, $N \cos \xi$, is also very similar compared to the results in the fully radiative regime. We therefore forgo to discuss the isothermal results in detail at this point.

\begin{figure}
 \centering
 \includegraphics[width=0.8\linwx]{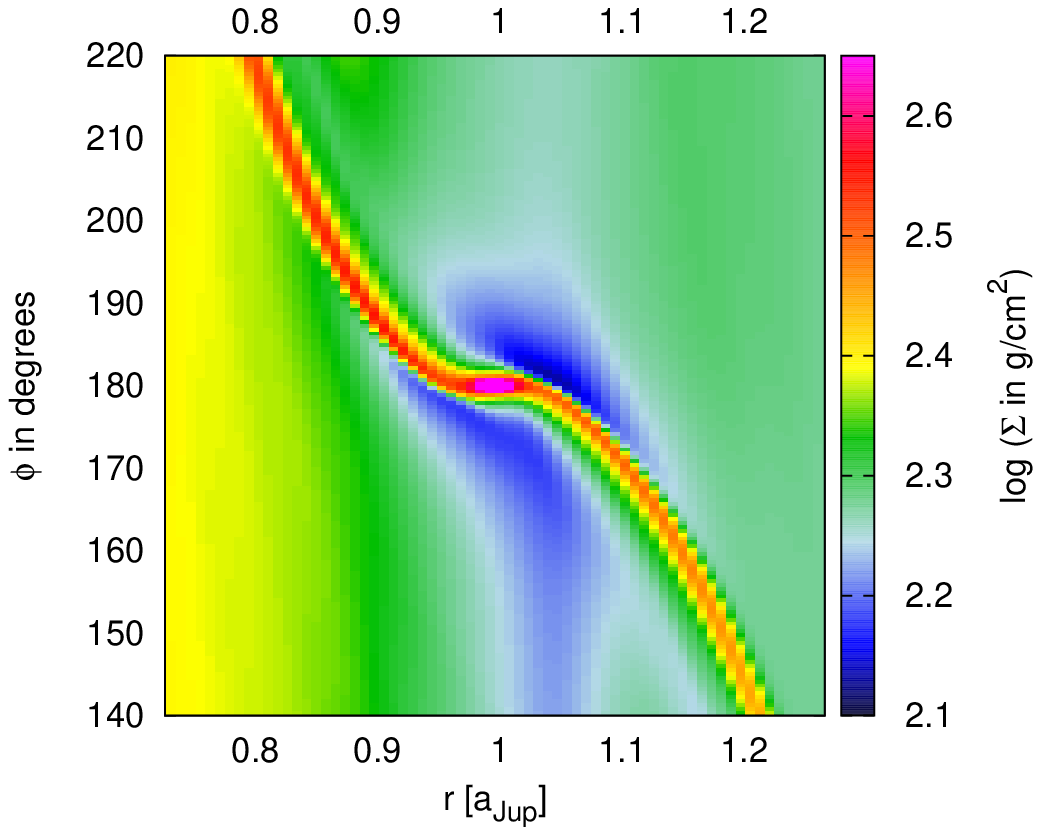}
 \includegraphics[width=0.8\linwx]{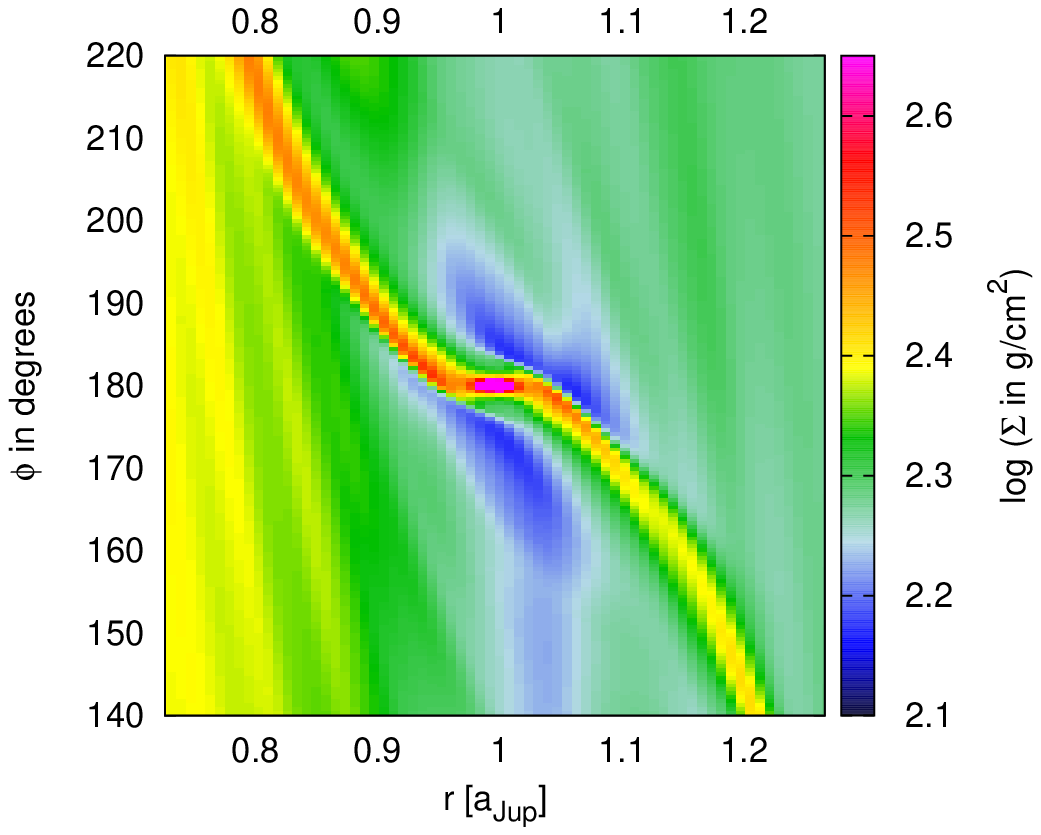}
 \caption{Surface density for $20 M_{Earth}$ planets on fixed circular inclined orbits embedded in an isothermal $H/r=0.037$ disc with $i=1.0^\circ$ (top) and $i=4.0^\circ$ (bottom). The planet is at its lowest point in orbit (lower culmination) at the time of the snapshot.
   \label{fig:IsoIncRhoxz}
   }
\end{figure}

To illustrate the dynamical impact of the planet on the disc, we display the surface density of planets with $i=1.0^\circ$ and $i=4.0^\circ$ in Fig.~\ref{fig:IsoIncRhoxz}.

The surface density of the $i=1.0^\circ$ planet is very similar to that of a non inclined planet (not displayed here, but compared to bottom picture of Fig.8 in \citet{2009A&A...506..971K}), and the torque density $\Gamma (r)$ is nearly identical. The surface density also shows no disturbances for the $i=1.0^\circ$ planet, so that we only observe the Lindblad Torques in the isothermal case, see Fig.~\ref{fig:IncGammaIsofull}. However, the surface density of the higher inclined planet ($i=4.0^\circ$) shows density disturbances near the planet. For the $i=1.0^\circ$ we also observe a higher density in the planets surroundings and a stronger pronounced spiral wave compared to higher inclined planets. The less pronounced spiral wave for the higher inclined planets leads to a smaller Lindblad Torque for these planets compared to the low inclined planets, which can be seen clearly in Fig.~\ref{fig:IncGammaIsofull}.

\begin{figure}
 \centering
 \includegraphics[width=1.0\linwx]{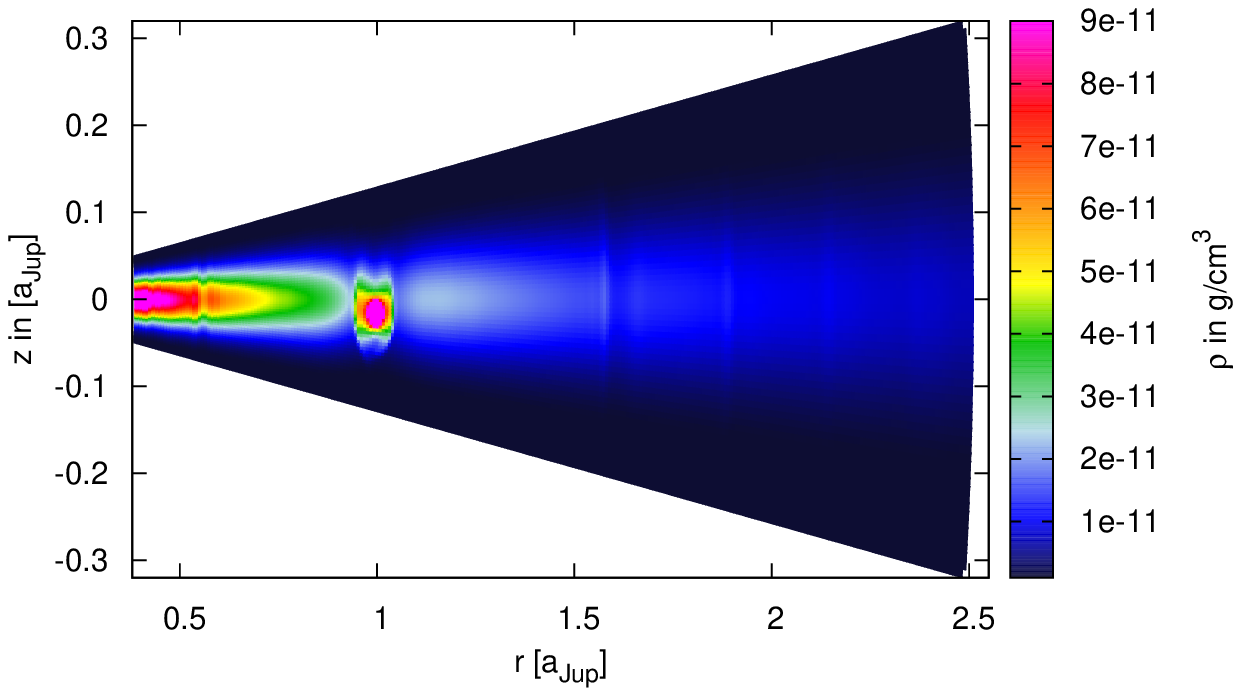}
 \includegraphics[width=1.0\linwx]{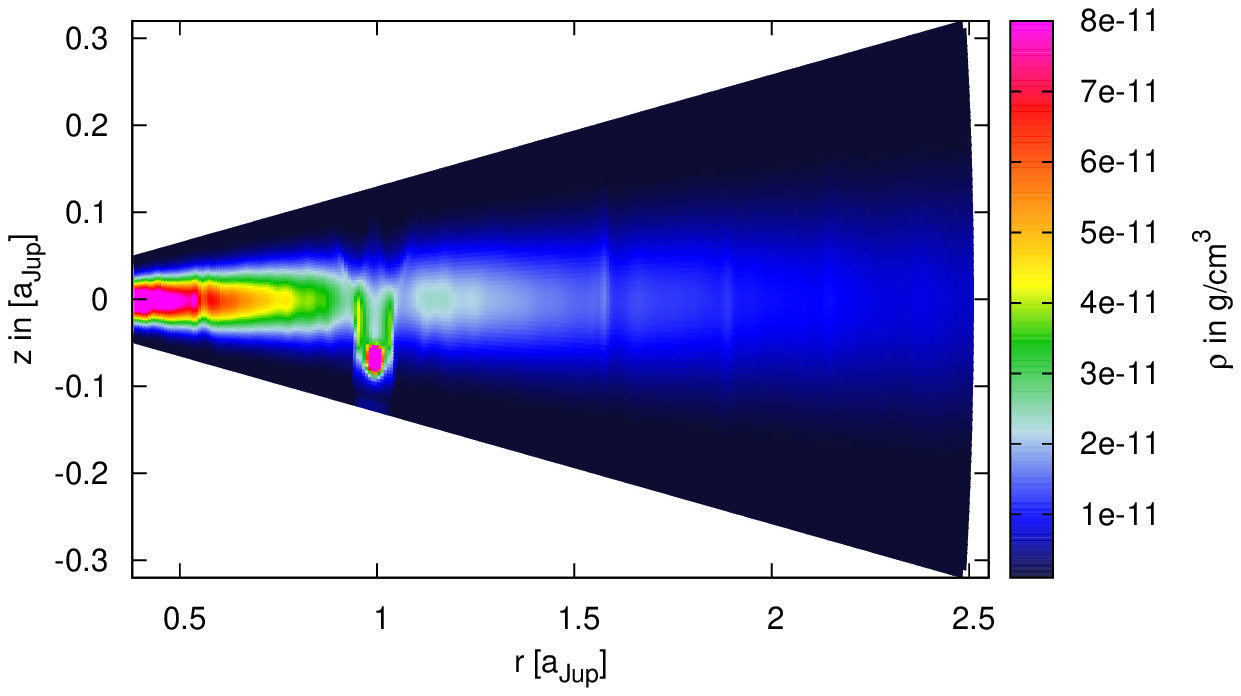}
 \caption{Density (in $g/cm^3$) of a $r-\theta$-slice through the isothermal $H/r=0.037$ disc with embedded $20 M_{Earth}$ planets on fixed circular inclined orbits with $i=1.0^\circ$ (top) and $i=4.0^\circ$ (bottom). The planet is at its lowest point in orbit (lower culmination) at the time of the snapshot.
   \label{fig:IsoIncRhoxy}
   }
\end{figure}

In Fig.~\ref{fig:IsoIncRhoxy} the density of a $r$-$\theta$ slice through the planets in the isothermal $H/r=0.037$ disc is display. The planets are at their lowest point in orbit when the snapshot was taken. On the one hand, the $i=1.0^\circ$ planet (top) has accumulated more mass in its vicinity as the $i=4.0^\circ$ planet (bottom), as the density in the middle of the disc is higher and thus it is easier to accumulate mass. On the other hand, the higher inclined planet seems to disturb the density structure of the disc stronger than the lower inclined planet. This distortion in the density distribution in the $r$-$\theta$ plane reflects in the distorted surface density structure, displayed in the bottom picture of Fig.~\ref{fig:IsoIncRhoxz}. 

In the fully radiative case, the torque density (Fig.~\ref{fig:IncGammaIsofull}) shows the well discussed spike in the torque distribution at $r=0.984$ (see \citet{2009A&A...506..971K}). This spike is more pronounced for lower inclinations ($i \leq 2.0^\circ$) and is reduced for the higher inclinations. This reduction for the torque density at $r=0.984$ causes the total torque to decrease for higher inclinations, see Fig.~\ref{fig:InctorIsofull}. For higher inclinations the Lindblad Torques are also reduced compared to the lower inclined planets.

\begin{figure}
 \centering
 \includegraphics[width=0.8\linwx]{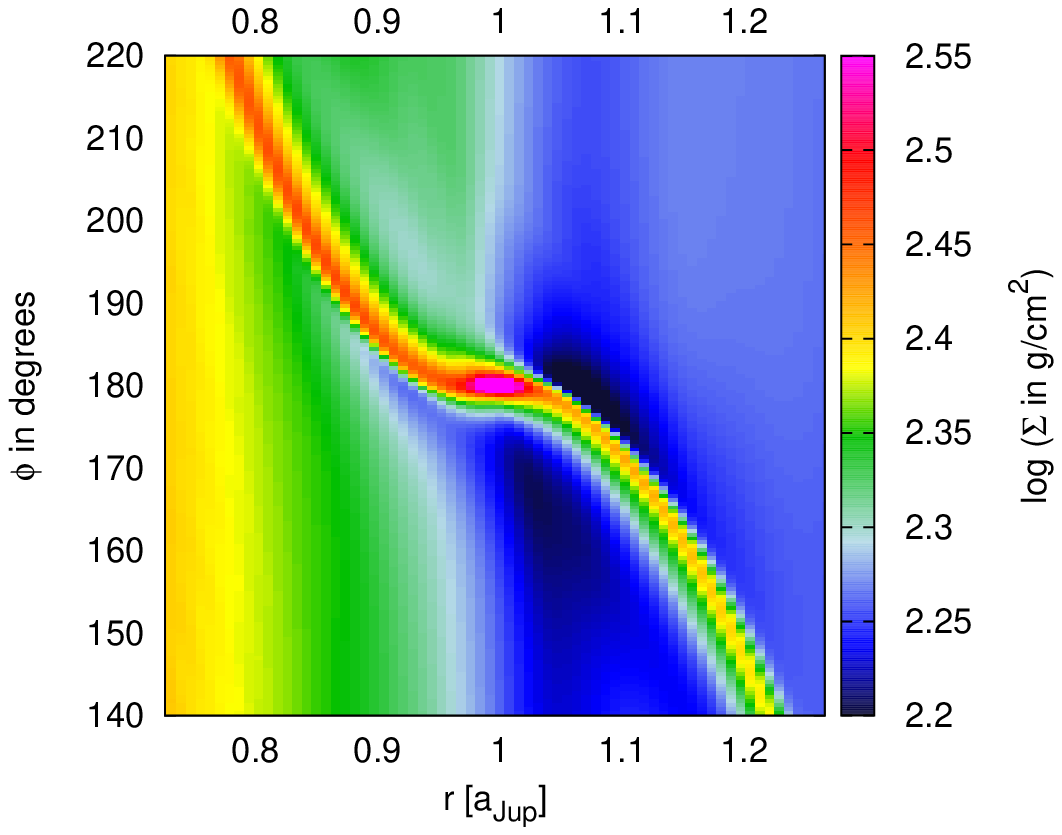}
 \includegraphics[width=0.8\linwx]{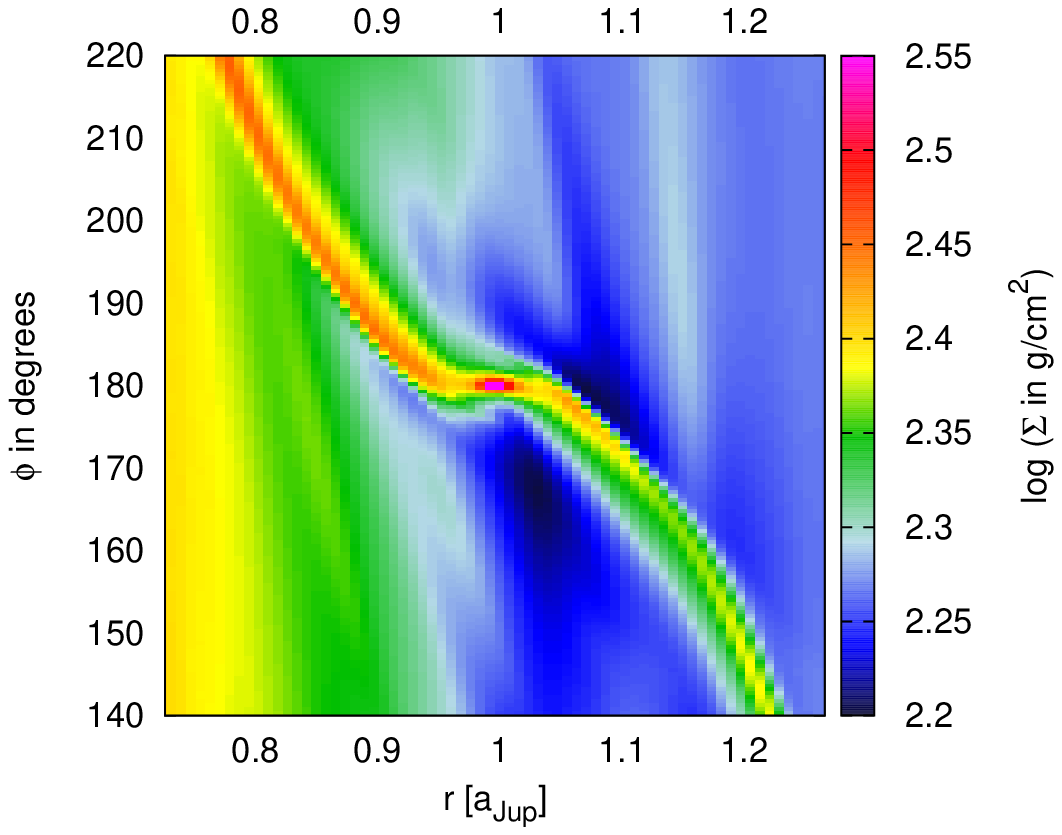}
 \caption{Surface density for $20 M_{Earth}$ planets on fixed circular inclined orbits embedded in a fully radiative disc with $i=1.0^\circ$ (top) and $i=4.0^\circ$ (bottom). The planet is at its lowest point in orbit (lower culmination) at the time of the snapshot.
   \label{fig:fullIncRhoxz}
   }
\end{figure}

In Fig.~\ref{fig:fullIncRhoxz} the surface density of $20 M_{Earth}$ planets in a fully radiative disc with $i=1.0^\circ$ and $i=4.0^\circ$ are displayed. The overall surface density distribution of the $i=1.0^\circ$ planet is nearly identical with the surface density distribution of a non inclined planet (see Fig.12 in \citet{2009A&A...506..971K}), which is also supported by the identical torque density distributions. The surface density for the higher inclined planet ($i=4.0^\circ$) on the other hand shows some differences. The density ahead of the planet ($r<1.0$ and $\phi>180^\circ$) is reduced and the density behind the planet ($r>1.0$ and $\phi<180^\circ$) is increased compared to the density distribution of the low inclined ($i=1.0^\circ$) planet. This loss and gain of density results directly in a reduced spike (at $r=0.984$) in the torque density (Fig.~\ref{fig:IncGammaIsofull}), which reduces the total torque. The density inside the planet's Roche lobe is also reduced in the high inclined case, as the planet can not accumulate as much mass as a lower inclined planet. This feature was also visible in the isothermal case, but keep in mind that the planet in the isothermal case accumulates more mass than in the fully radiative disc. Also the density in the spiral waves is reduced in the high inclined case, which results in lower Lindblad Torques compared to the low inclined case, see Fig.~\ref{fig:IncGammaIsofull}.

\begin{figure}
 \centering
 \includegraphics[width=1.0\linwx]{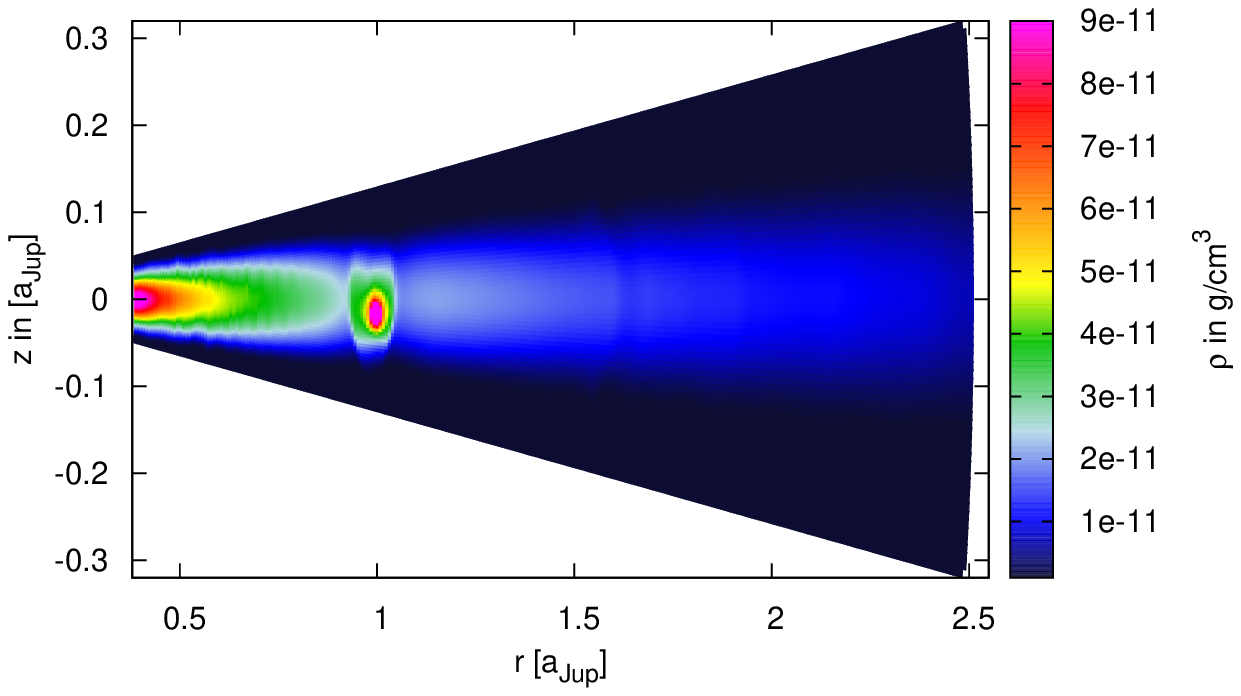}
 \includegraphics[width=1.0\linwx]{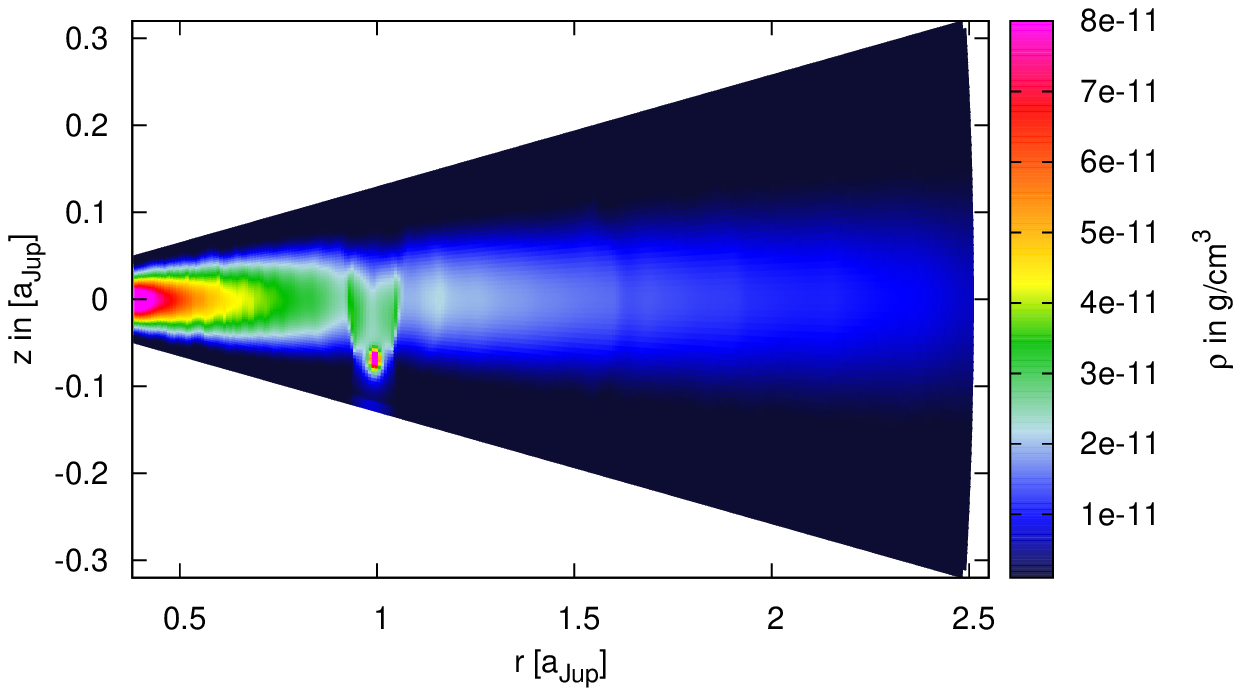}
 \caption{Density (in $g/cm^3$) of a $r-\theta$-slice through the fully radiative disc with embedded $20 M_{Earth}$ planets on fixed circular inclined orbits with $i=1.0^\circ$ (top) and $i=4.0^\circ$ (bottom). The planet is at its lowest point in orbit (lower culmination) at the time of the snapshot.
   \label{fig:fullIncRhoxy}
   }
\end{figure}

In Fig.~\ref{fig:fullIncRhoxy} the density through the fully radiative disc (a $r$-$\theta$ slice) at the planets location for $i=1.0^\circ$ and $i=4.0^\circ$ is displayed. Clearly the higher inclined planet accumulated less mass than the lower inclined planet, as the density in the surroundings of the high inclined planet is reduced compared to the discs midplane. Also the higher inclined planet seems to disturb the disc's density more than the lower inclined planet, which results in small fluctuations in the surface density, see Fig.~\ref{fig:fullIncRhoxz}. The little fluctuations in the surface density are less compared to the isothermal case, as the radiative transport/cooling smoothes the density in the disc and the planet does not accumulate as much mass as in the isothermal case, which disturbs the density structure as well.

In the $r-\theta$ slices of the disc, the difference in the mass accumulation of the planet in the isothermal and fully radiative regime is very obvious. This effect is caused by the thermodynamics of the disc, as the isothermal disc is unable to heat and cool, the planet can accumulate more mass as in the fully radiative regime. Nevertheless, only in the fully radiative regime the calculated change of semi-major axis predicts outward migration for planets on fixed orbits with $i_0 \leq 4.5^\circ$.

\section{Moving planets on initially inclined orbits}
\label{sec:moveplanets}

In the previous section we calculated the change of inclination and semi-major axis for $20 M_{Earth}$ planets on fixed inclined orbits in the isothermal and fully radiative regime. The results stated that the planets will lose their inclination in time and will migrate inwards in the isothermal and outwards in the fully radiative scheme (for $i_0 \leq 4.5^\circ$). We now want to confirm these results by allowing the planets to move freely inside the disc. For the following simulations we use our standard resolution, with a $20 M_{Earth}$ planet embedded at $r=1.0$ with different inclinations $i$. The discs thickness is $H/r=0.037$. In the first $10$ orbits of the planetary evolution the mass of the planet will rise until it reaches its desired mass of $20 M_{Earth}$ after $10$ planetary orbits. This way the disc will not be disturbed as much as by putting a planet with its full mass inside the disc at once. By monitoring inclination and semi-major axis at the same time for a given planet, we can easily observe what influence the inclination has on the migration of the planet. The planets embedded in the isothermal disc are modelled with the $\epsilon$-potential, using $r_{sm}=0.8$, while the planets embedded in the fully radiative disc are modelled using the cubic potential with $r_{sm}=0.5$.

\subsection{Isothermal disc}

Our disc only extends $7^\circ$ above and below the discs midplane, so when simulating a planet with a higher inclination than that, it is not fully embedded in the disc any more. As the disc's material is concentrated mainly in the middle of the disc, the material high above or below the disc's midplane will only have a small effect on the planet anyway. Test calculations with an extended $\theta$ region of the disc ($10^\circ$ above and below the discs midplane) and an $i_0=8.0^\circ$ $20 M_{Earth}$ planet have shown that we do not need such a high extension above and below the disc. The aspect ratio for the disc is $H/r=0.037$ in the isothermal case, as this aspect ratio corresponds to the aspect ratio of the fully radiative disc. 

\begin{figure}
 \centering
 \includegraphics[width=0.9\linwx]{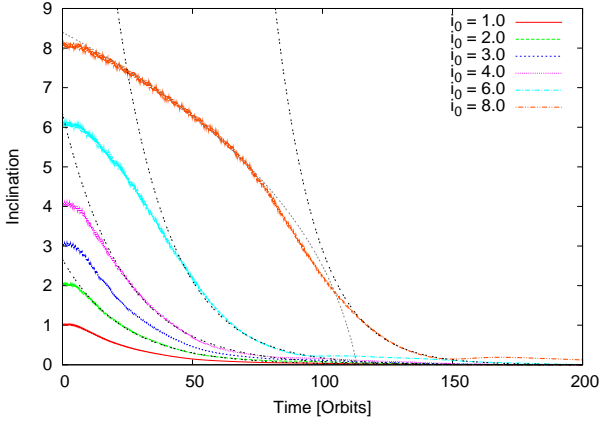}
 \caption{Time evolution of inclination for $20 M_{Earth}$ planets with individual starting inclinations ranging from $i=1.0^\circ$ to $i=8.0^\circ$ in an isothermal $H/r=0.037$ disc. The damping of inclination follows the trend of the calculated change of inclination for planets on fixed orbits, see Fig.~\ref{fig:Incchangeplot}. The black dotted lines indicate our manual fitting with $\tau_{inc}=23$ and in the $i_0=8.0^\circ$ case an additional fit provides for a $di/dt \propto i^{-2}$ fit.
   \label{fig:IncIncIso}
   }
\end{figure}

In Fig.~\ref{fig:IncIncIso} the evolution of inclination of a $20 M_{Earth}$ planet in an isothermal $H/r=0.037$ disc is displayed. After the planets have reached their final mass after $10$ orbits, the inclination begins to drop immediately. Up to $i=4.0^\circ$ the planet's loss of inclination is increasing, while for even higher inclinations the damping of inclination is slowed down again (see Fig.~\ref{fig:Incchangeplot}). High inclined planets will first lose inclination at a quite slow rate, but as their inclination is damped the damping rate increases until the inclination reaches $\approx 4.0^\circ$ and will slow down after that until the planets inclination is damped to zero. This results confirms our results for planets on fixed orbits.

When the planet remains in the main body of the disc for $i < H/r$ ($i$ in radians), the damping of inclination is exponential: $di/dt \propto -i$. By following \citet{2004ApJ...602..388T} the linear damping rate for a small planet mass and a small inclination can be obtained:
\begin{equation}
\label{eq:idot}
 \frac{\overline{di/dt}}{i}= - \frac{1}{\tau_{inc}} = - \frac{0.544}{t_{wave}}
\end{equation}
with the characteristic time
\begin{equation}
\label{eq:twave}
	t_{wave}=q^{-1} \left(\frac{\Sigma_P a^2}{M_\ast}\right)^{-1} \left(\frac{c_s}{a\Omega_P}\right)^4 \Omega_P^{-1} \ .
\end{equation}

In Tab.~\ref{tab:tauinc} we display $\tau_{inc}$ for the isothermal and fully radiative simulations. We compare the linear estimate according to eqs.~\ref{eq:idot} and
\ref{eq:twave} with our full non-linear results from the fixed and moving planet simulations.

{%
\newcommand{\mc}[3]{\multicolumn{#1}{#2}{#3}}
\begin{table}[b]
 \centering
 \begin{tabular}{l|l|l|l|}\cline{2-4}
  & \textbf{linear 3D} & \textbf{fixed orbit} & \textbf{moving orbit}\\\hline
 \mc{1}{|l|}{\textbf{iso $H/r=0.037$}} & 14.2 & 27 & 23\\\hline
 \mc{1}{|l|}{\textbf{iso $H/r=0.05$}} & 47.25 & $-$ & 53\\\hline
 \mc{1}{|l|}{\textbf{fully radiative}} & 27.7 & 20 & 26.5\\\hline
 \end{tabular}
 \caption{The exponential damping rate $\tau_{inc}$ (in orbits), obtained from our isothermal and fully radiative simulations for three studied cases of planets in discs compared with the linear rate, as given by \citet{2004ApJ...602..388T}. 
Shown are results for planets on fixed orbits and for moving planets.
 \label{tab:tauinc}
 }
\end{table}
}%

For our isothermal $H/r=0.037$ simulations we estimate a linear damping time scale of $\tau_{inc} = 14.2$ orbits, which is about a third smaller than our obtained result $\tau_{inc} = 23$ for planets on moving orbits (see fits in Fig.~\ref{fig:IncIncIso}). The difference between our actual fit and the linear value can be caused by the relatively small aspect ratio of the disc ($H/r=0.037$), because the linear theory is formally valid only for embedded planets with $H \gg R_{Hill}$. Below, we present additional result for an $H/r=0.05$ disc and find indeed better agreement of linear and fully non-linear results. However, our obtained $\tau_{inc}=23$ for moving planets is very close to the predicted damping time scale of $\tau_{inc}=27$ for planets on fixed orbits, demonstrating the consistency of our results.

The exponential damping law should be only valid up to $i \approx H/r$, but \citet{2007A&A...473..329C} stated that it is valid up to $i \approx 2 H/r$, which in fact reflects our findings. The higher inclined planets do not lose inclination with an exponential rate at the beginning of the simulations, but with $di/dt \propto i^{-2}$. This is also indicated through the black dotted line in Fig.~\ref{fig:IncIncIso} for the $i_0=8.0^\circ$ planet. As the planets lose their inclination, the damping becomes exponential again (at about $i \approx 3.0^\circ$).

\begin{figure}
 \centering
 \includegraphics[width=0.9\linwx]{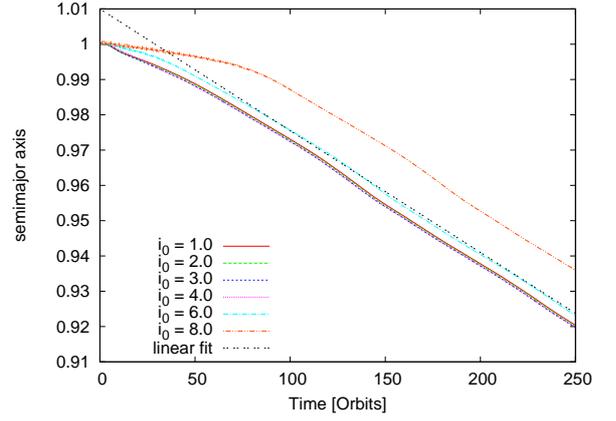}
 \caption{Time evolution of semi-major axis for $20 M_{earth}$ planets with individual starting inclinations ranging from $i=1.0^\circ$ to $i=8.0^\circ$ in an isothermal $H/r=0.037$ disc. 
 The dashed black line indicates a fit to the results with a slope of $- 3.45 \, 10^{-4}$/orbit.
  The change of semi-major axis follows the trend of the change {determined for planets on fixed orbits}, see Fig.~\ref{fig:AIncIsofullrate}.
   \label{fig:AIncIso}
   }
\end{figure}

In Fig.~\ref{fig:AIncIso} the evolution of the semi-major axis in an isothermal $H/r=0.037$ disc is displayed. After a few orbits the semi-major axis decreases for all planets. During time the semi-major axis shrinks more and more, as you would expect from low mass planets in an isothermal disc. In the beginning the planets with $i_0=8.0^\circ$ and $i_0=6.0^\circ$ have a slower inward migration rate than the other planets, but as the inclination is damped in time, the inward migration settles for the same rate for all planets, although it takes about $80$ orbits for the $i_0=8.0^\circ$ planet. The initial higher inclination just delays the inward migration for a few planetary orbits, as could be expected from our results for planets on fixed orbits, see Fig.~\ref{fig:AIncIsofullrate}.

The observed migration rate for all planets is constant after the inclination has been damped. The black dashed line in Fig.~\ref{fig:AIncIso} indicates a linear fit for the evolution of the semi-major axis with $\dot{a} = - 3.45 \, 10^{-4}$ / orbit. This decrease of semi-major axis from planets on moving orbits is within 15\% of the result for planets on fixed orbits in Fig.~\ref{fig:AIncIsofullrate}. The slight difference may be just due to the different setup, stationary versus moving and cubic versus epsilon potential. Additionally, the fit in Fig.~\ref{fig:AIncIso} is for a later evolutionary time, where the planet has already lost a few percent of its semi-major axis.

\begin{figure}
 \centering
 \includegraphics[width=0.9\linwx]{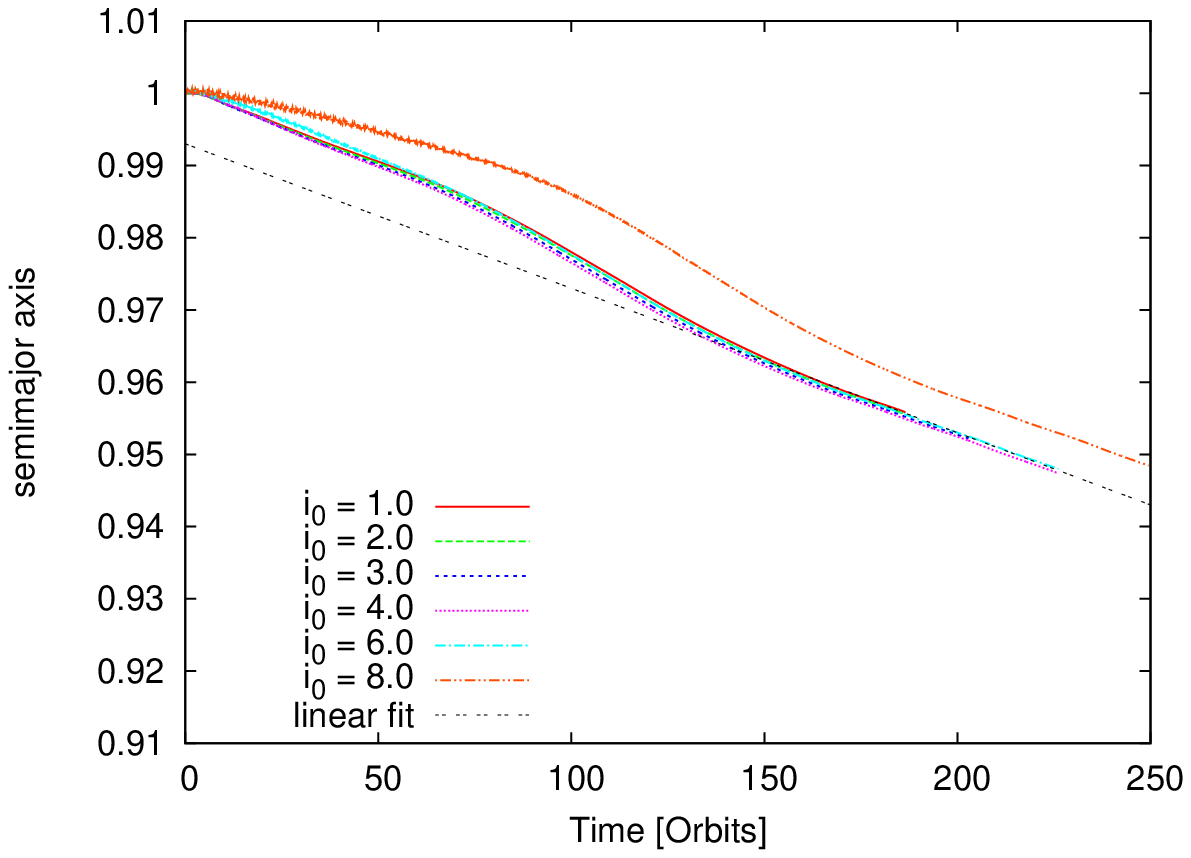}
 \includegraphics[width=0.9\linwx]{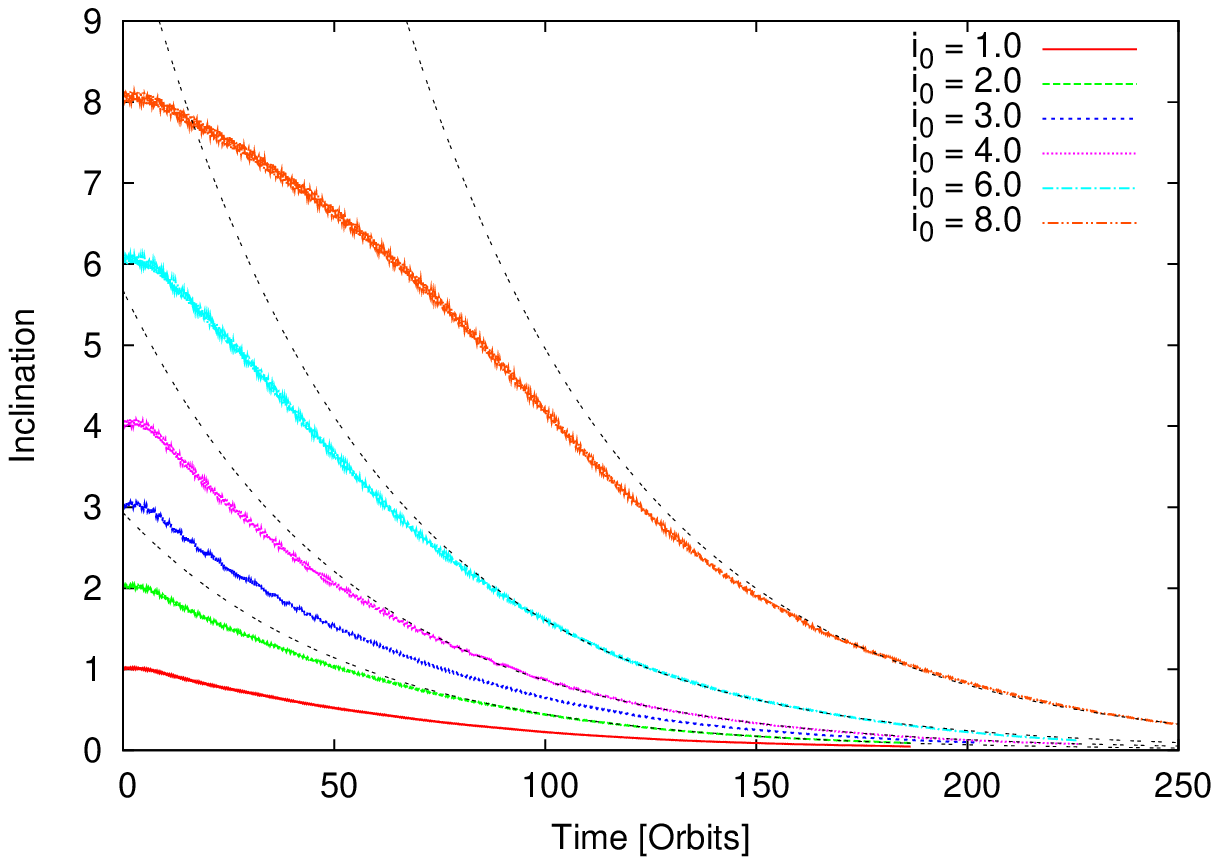}
 \caption{Time evolution of the semi-major axis (top) and inclination (bottom) of a $20 M_{Earth}$ planet in an isothermal $H/r=0.05$ disc. The planets initial inclinations reach from $i_0=1.0^\circ$ to $i_0=8.0^\circ$. The dashed black line in the top panel indicates a linear fit to the results with a slope of $- 2 \, 10^{-4}$/orbit. The black dotted lines in the evolution of inclination (bottom) indicate our manual fitting with $\tau_{inc}=53$.
   \label{fig:AIncIsoHr005}
   }
\end{figure}

In our past simulations we found a dependence of the migration rate/torques due to the aspect ratio of the disc \citep{2009A&A...506..971K,2010A&A.523...A30}. It is now logical to assume that the aspect ratio of a disc will change the damping of inclination for inclined planets as well. In Fig.~\ref{fig:AIncIsoHr005} we display the evolution of the semi-major axis (top) and inclination (bottom) of planets embedded in an isothermal $H/r=0.05$ disc. 

The inward migration of the planets in the $H/r=0.05$ disc is slower compared to planets embedded in a $H/r=0.037$ disc, which is in agreement with \citet{2002ApJ...565.1257T}. The inclination damping is slower in the $H/r=0.05$ disc by a factor of $2$ to $3$. Nevertheless, the final outcome is the same, an initially high inclination reduces the inward migration in the beginning, but as inclination is damped, the inward migration settles for the same rate for all the inclinations (but with different rates for different aspect ratios).

The theoretical, linear damping of inclination obtained by following \citet{2004ApJ...602..388T} is $\tau_{inc} = 47.25$ orbits (eq.~\ref{eq:twave}), which is slightly smaller than our fitted value of $\tau_{inc}=53$ orbits (Fig.~\ref{fig:AIncIsoHr005}, bottom). The numerically obtained ratio of the migration rates ($53/23\approx 2.3$) differs from the theoretical linear ratio of $(0.05/0.037)^4 \approx 3.3$. As the colder disc makes linear
theory less applicable in these circumstances, we can infer that the results obtained for the larger $H/r = 0.05$ disc are more accurate.
The numerical exponential decay in the simulations with moving planets matches our estimated rate for planets on fixed orbits quite well. The higher pressure inside the disc seems to have a smoothing effect on the damping, resulting in more accurate results for discs with higher aspect ratio. 

For the moving planets, the measured migration rate ($da/dt$) for the $H/r=0.037$ disc is approximately larger by a factor of $(0.05/0.037)^2$ compared to the $H/r=0.05$ disc (see dashed black line fit in Fig.~\ref{fig:AIncIsoHr005}, top), which is to be expected in the linear case.

In an isothermal disc, inclined planets lose inclination and semi-major axis immediately after they are released in the disc, as predicted by our calculations of the change of inclination and semi-major axis for planets on fixed orbits. The migration rate is the same for all initial inclinations, a higher inclination just delays the inward migration for a few orbits.

\subsection{Fully radiative disc}

In the fully radiative regime we simulate planets to an inclination up to $i=8.0^\circ$. 
As always we use a $2D$ model in $r$-$\theta$ direction in the radiative equilibrium for the starting configuration of our $3D$ fully radiative disc. This procedure is described in more detail in \citep{2009A&A...506..971K}. As in the isothermal case, the planet needs $10$ planetary orbits to reach its full mass.

\begin{figure}
 \centering
 \includegraphics[width=0.9\linwx]{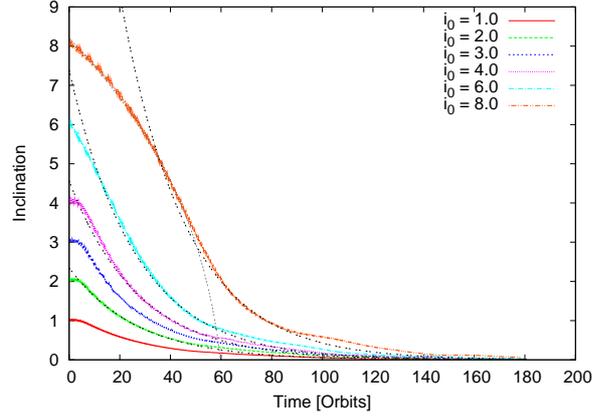}
 \caption{Time evolution of inclination for $20 M_{Earth}$ planets with individual starting inclinations ranging from $i=1.0^\circ$ to $i=4.0^\circ$ in a fully radiative disc. The damping of inclination follows the trend of change of inclination for planets on fixed orbits, see Fig.~\ref{fig:Incchangeplot}. The black dotted lines indicate our manual fitting with $\tau_{inc}=26.5$ for planets with $i_0=2.0^\circ, 4.0^\circ, 6.0^\circ$ and $i_0=8.0^\circ$, which seems very good in the beginning/middle phase of the inclination damping. In the $i_0=8.0^\circ$ case an additional fit provides for a $di/dt \propto i^{-2}$ fit in the beginning of the evolution.
   \label{fig:IncIncfull}
   }
\end{figure}

In Fig.~\ref{fig:IncIncfull} the change of inclination for moving planets in a fully radiative disc are displayed. The inclinations range from $i_0=1.0^\circ$ to $i_0=8.0^\circ$. The planets inclination reduces as soon as the planet is released in the disc according to the rate found in Fig.~\ref{fig:Incchangeplot}. After about $160$ orbits the inclination of all planets has reached zero, so that the planet is now orbiting in the equatorial plane.

The estimated linear value for the damping of inclination in the radiative disc is $\tau_{inc}=27.76$ orbits (eq.~\ref{eq:twave}), which lies within about a few percent of our fitted value of $\tau_{inc}=26.5$ orbits for moving planets. The difference with the calculated damping timescale $\tau_{inc}=20$ orbits for planets on fixed orbits is slightly larger in this case (see Tab.~\ref{tab:tauinc}). But one should keep in mind as well that the exponential fit is valid for very low inclinations in the isothermal case as well, while it is only valid down to $\approx 0.5^\circ$ in the fully radiative scheme. 

On the other hand, the decay of inclination for initially high inclined planets fits even a little better than in the isothermal case, and the trend of an $di/dt \propto i^{-2}$ decay is clearly visible. 

\begin{figure}
 \centering
 \includegraphics[width=0.9\linwx]{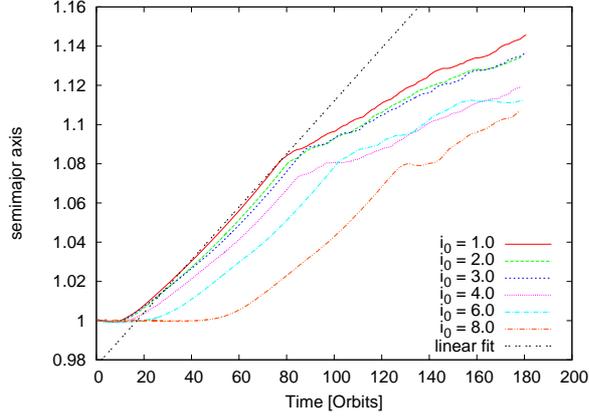}
 \caption{Time evolution of semi-major axis for $20 M_{Earth}$ planets with individual starting inclinations ranging from $i=1.0^\circ$ to $i=8.0^\circ$ in a fully radiative disc. The change of semi-major axis follows the trend of change for planets on fixed orbits, see Fig.~\ref{fig:AIncIsofullrate}. The dashed black line in the top panel indicates a linear fit to the results with a slope of $ 1.35 \, 10^{-3}$/orbit.
   \label{fig:AIncfull}
   }
\end{figure}

In Fig.~\ref{fig:AIncfull} the evolution of the semi-major axis of planets with initial inclination is displayed. After the planets have reached their final mass at $10$ planetary orbits, the initially lower inclined ($i$ up to $4.0^\circ$) planets start to migrate outwards as expected from Fig.~\ref{fig:AIncIsofullrate}. The migration rates for planets on fixed orbits also matches nicely with the determined migration rate of $ 1.35 \, 10^{-3}$/orbit for planets on moving orbits. The two higher inclined planets first stay on their initial orbit before they start their outward migration. It seems that the planet has to lose a certain amount of inclination before it can start to migrate outwards. This result is agreement with our observations for planets on fixed inclined orbits. A higher inclination weakens the torque responsible for outward migration and if the inclination is too high, the outward migration stops.

The planets seem to migrate outwards at nearly the same speed for all initial inclinations up to $4.0^\circ$. But before the planets start to migrate outwards their initial inclination is damped by about $25\%$ for all initial inclinations, which explains the similar migration speed for these planets, see Fig.~\ref{fig:AIncIsofullrate}. 

Low inclinations seem to have only little effect on the migration of planets in the fully radiative scheme, but if the inclination of the planet is so high that its orbit is at some times high above the midplane of the disc, the disc becomes so thin that outward migration is not possible any more.

\section{Planets with different masses}
\label{sec:massplanets}

Small inclinations seem to have no big effect on migration for a $20 M_{Earth}$ planet on a circular orbit in the isothermal or fully radiative scheme. The planet's inclination is reduced in both thermodynamic systems and it migrates inwards in the isothermal and outwards in the fully radiative scheme. In our previous work we found that planets up to $\approx 33 M_{Earth}$ migrate outwards in a fully radiative disc \citep{2009A&A...506..971K}. Planets with higher planetary masses migrate inwards, and considering our previous results it is unlikely that inclination changes this. Nevertheless we here present simulations of planets with masses ranging from $20$ up to $100 M_{Earth}$ in an isothermal $H/r=0.037$ disc and in a fully radiative disc. We only present results for moving planets with our usual tool for increasing the planetary mass to its designated mass during the first $10$ orbits.

For all simulations we limit ourselves to the cases of $i_0 = 1.0^\circ$ and $i_0 = 4.0^\circ$ for all planetary masses, in the isothermal $H/r=0.037$ disc as well as in the fully radiative disc. Additionally, we display Jovian mass planets with inclinations up to $15^\circ$ in the isothermal case. For the isothermal case, we again use the $\epsilon$-potential with a smoothing length of $r_{sm}=0.8$, as it turned out to be more stable in the isothermal case than the cubic potential, while we use the cubic $r_{sm}=0.5$ potential for the fully radiative simulations.

In all simulations displayed in this section the planet is inserted in the disc and starts to move immediately. This means that a massive planet will open a gap in time in the initially unperturbed disc. If the planet were allowed to open a gap before it starts to move in the disc, the evolution of the planet, especially the evolution of inclination, would be slower since an open gap slows down migration and inclination damping.

\subsection{Isothermal disc}

\begin{figure}
 \centering
 \includegraphics[width=0.9\linwx]{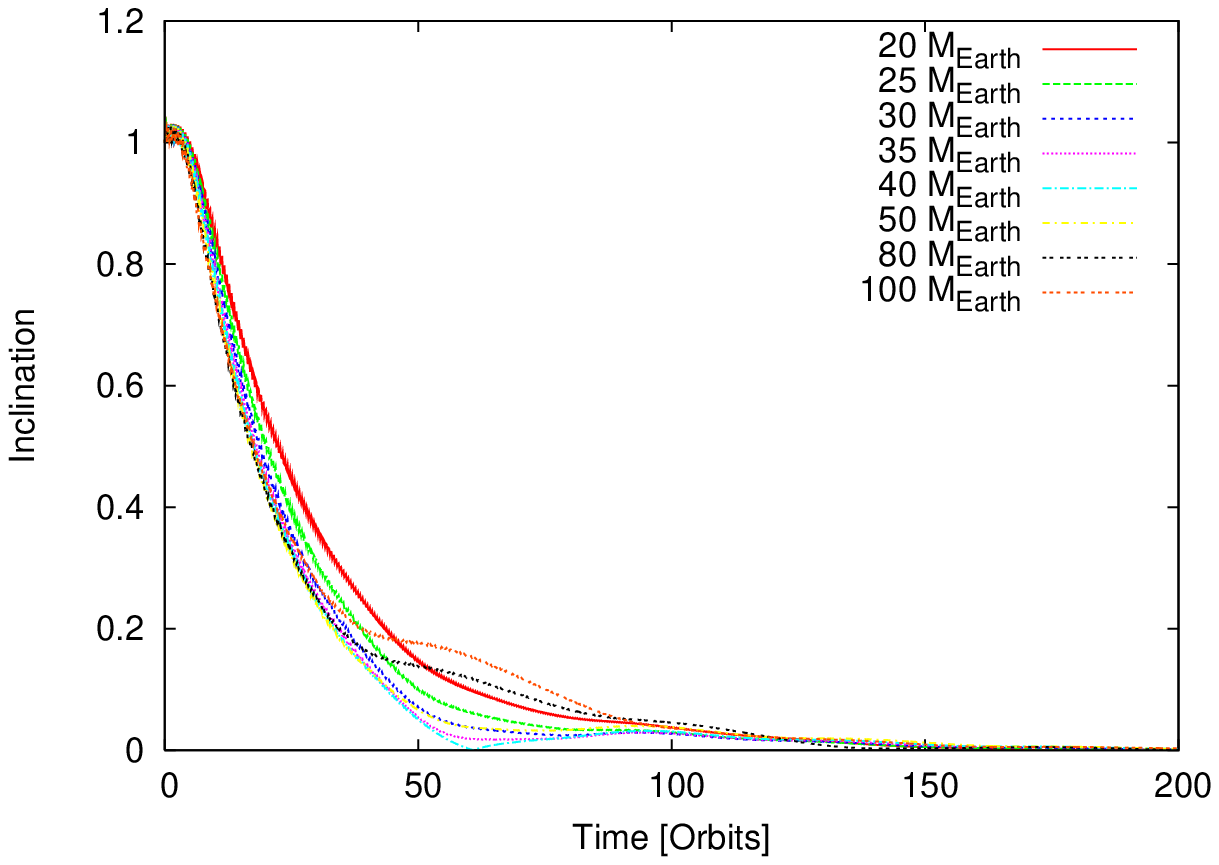}
 \includegraphics[width=0.9\linwx]{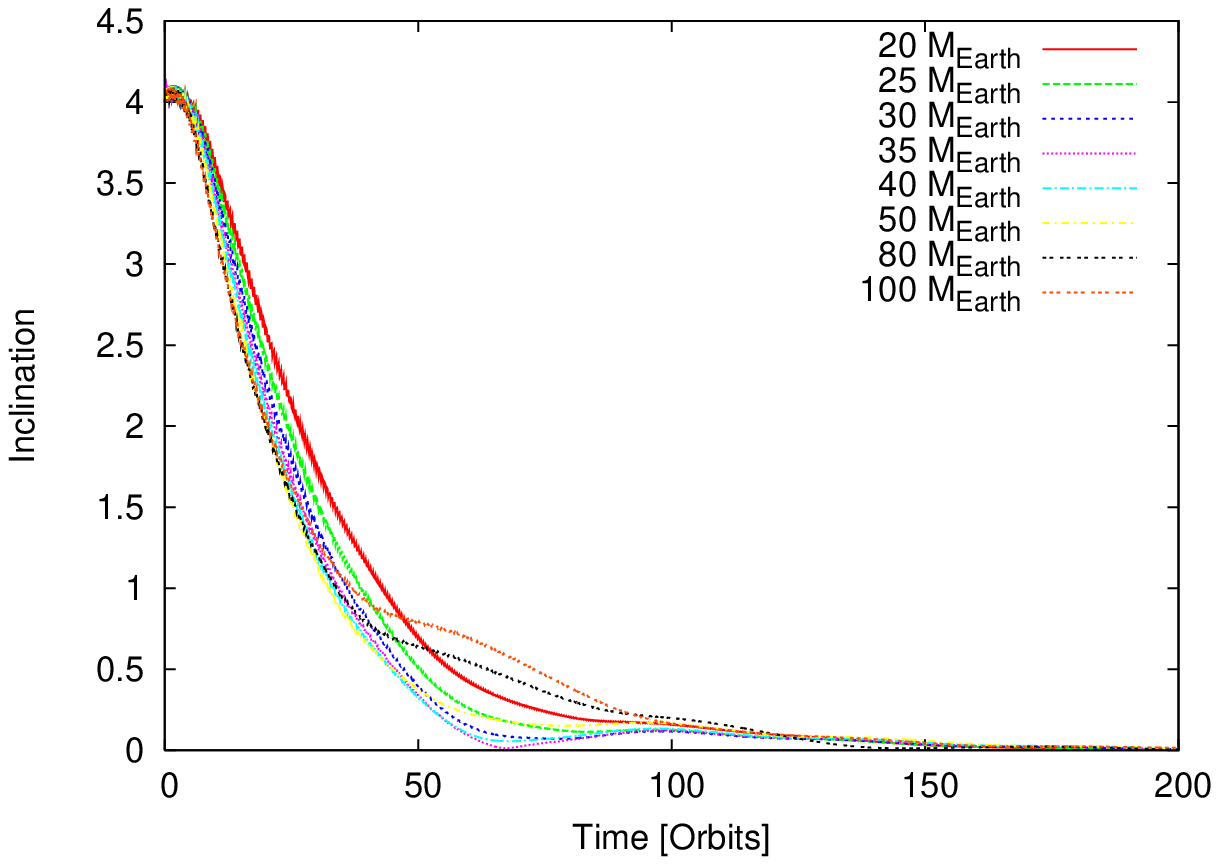}
 \caption{Time evolution of the inclination for planets with $i_0 = 1.0^\circ$ (top) and $i_0 = 4.0^\circ$ (bottom) for different planetary masses in an isothermal $H/r=0.037$ disc.
   \label{fig:IncInc1040Iso}
   }
\end{figure}

In Fig.~\ref{fig:IncInc1040Iso} the evolution of inclination for planets with $i_0 = 1.0^\circ$ (top) and $i_0 = 4.0^\circ$ (bottom) for different planetary masses in an isothermal $H/r=0.037$ disc is displayed. The inclination drops immediately after the planet reaches its destined mass at nearly the same rate for all planetary masses in both cases, for the $i_0 = 1.0^\circ$ and $i_0 = 4.0^\circ$ case. It seems that a higher planetary mass results in a slightly faster damping rate for the inclination. However, when the inclination reaches about $20\%$ of the initial inclination, the $80$ and $100 M_{Earth}$ planets stop their fast inclination damping for a few orbits. After about $150$ orbits the inclination of all planets is damped to zero, even for the initially more highly inclined planets.

\begin{figure}
 \centering
 \includegraphics[width=0.9\linwx]{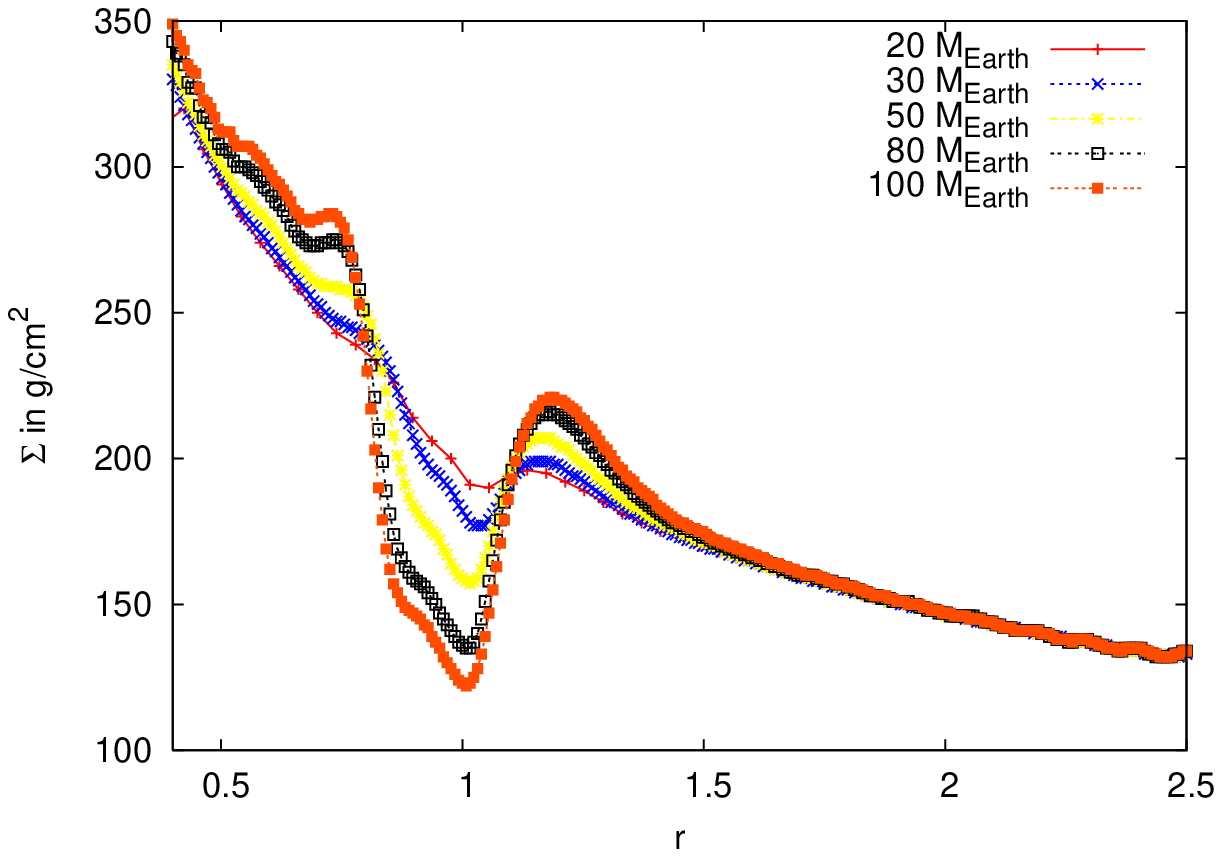}
 \includegraphics[width=0.9\linwx]{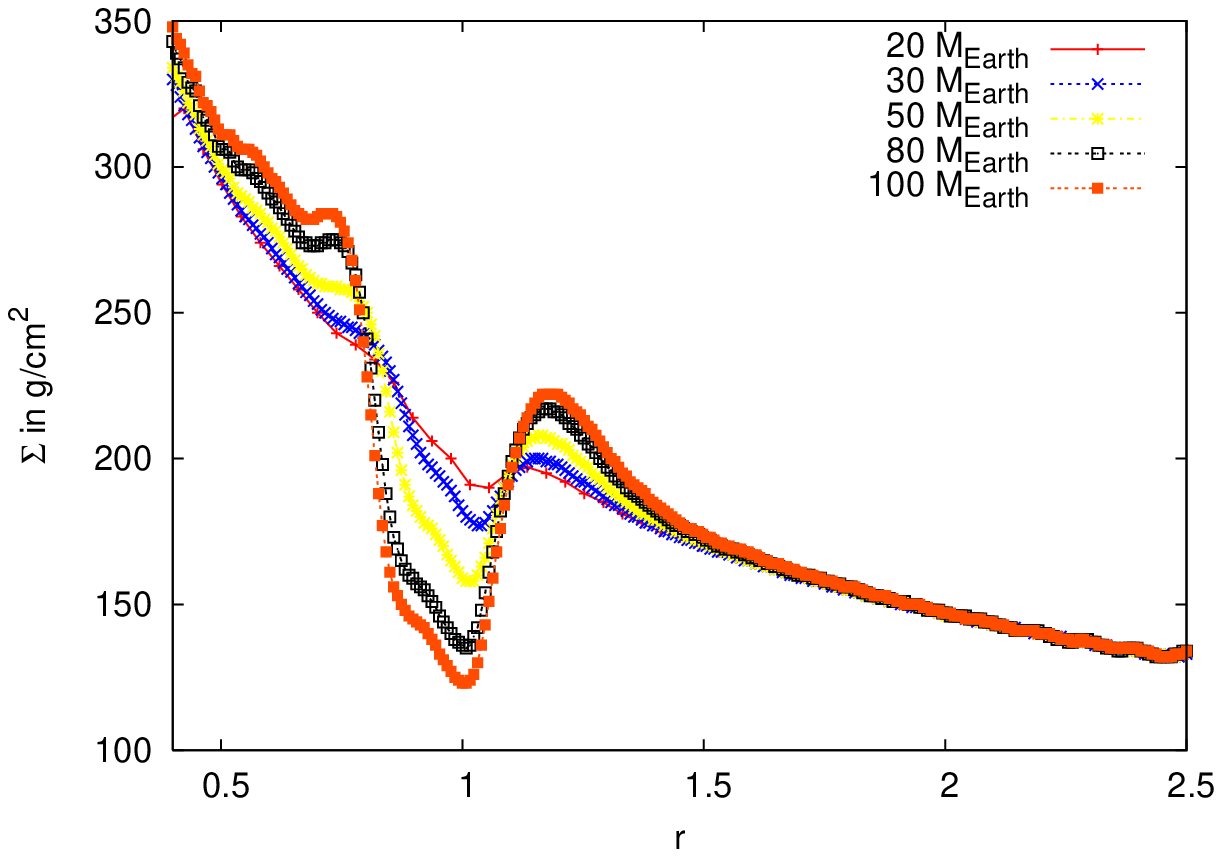}
 \caption{Surface density for planets on initial inclined, circular orbits with $i_0=1.0^\circ$ (top) and $i_0=4.0^\circ$ (bottom) after $60$ planetary orbits.
   \label{fig:Sig1040Iso}
   }
\end{figure}

In Fig.~\ref{fig:Sig1040Iso} we display the surface densities for planets with different masses for the $i_0=1.0^\circ$ (top) and $i_0=4.0^\circ$ (bottom) cases after $60$ planetary orbits. The two surface density profiles are very similar, although the planets have a different starting inclination $i_0$ and a different inclination at the time of the snapshot. As the inclination of the $i_0=4.0^\circ$ and $i_0=1.0^\circ$ planets evolves very similar (in the trend), it seems not surprising that the surface densities reflect this behaviour. On the other hand, the surface density profiles after $60$ orbits do not give a clear hint to explain the slowed down inclination damping for the high mass planets when they reach $\approx 20 \%$ of the initial inclination. As expected, higher mass planets clear deeper and wider gaps inside the disc.

\begin{figure}
 \centering
 \includegraphics[width=0.9\linwx]{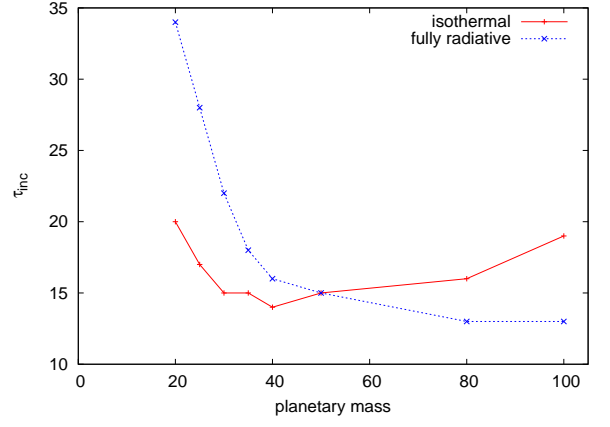}
 \caption{Measured inclination damping time scale $\tau_{inc}$ for planets with different planetary masses in the isothermal and fully radiative regime.
   \label{fig:tauincIsofull}
   }
\end{figure}

The measured damping time scale $\tau_{inc}$ for planets with different planetary masses is displayed in Fig.~\ref{fig:tauincIsofull}. Starting from the smallest planet in our calculations, the $20 M_{Earth}$, the damping time scale reduces until the planets mass is $40 M_{Earth}$. For higher mass planets the damping time scale increases again. The increase for the $80$ and $100 M_{Earth}$ planets should be handled with care, as the measurement for these planets is quite difficult due to the bump in the evolution of inclination. If we do not include these two planets in our thoughts, we can safely conclude, that the inclination damping increases with the planetary mass. The inclination of a more massive planet will be damped faster than the inclination of a smaller planet, even more so for radiative discs.

\begin{figure}
 \centering
 \includegraphics[width=0.9\linwx]{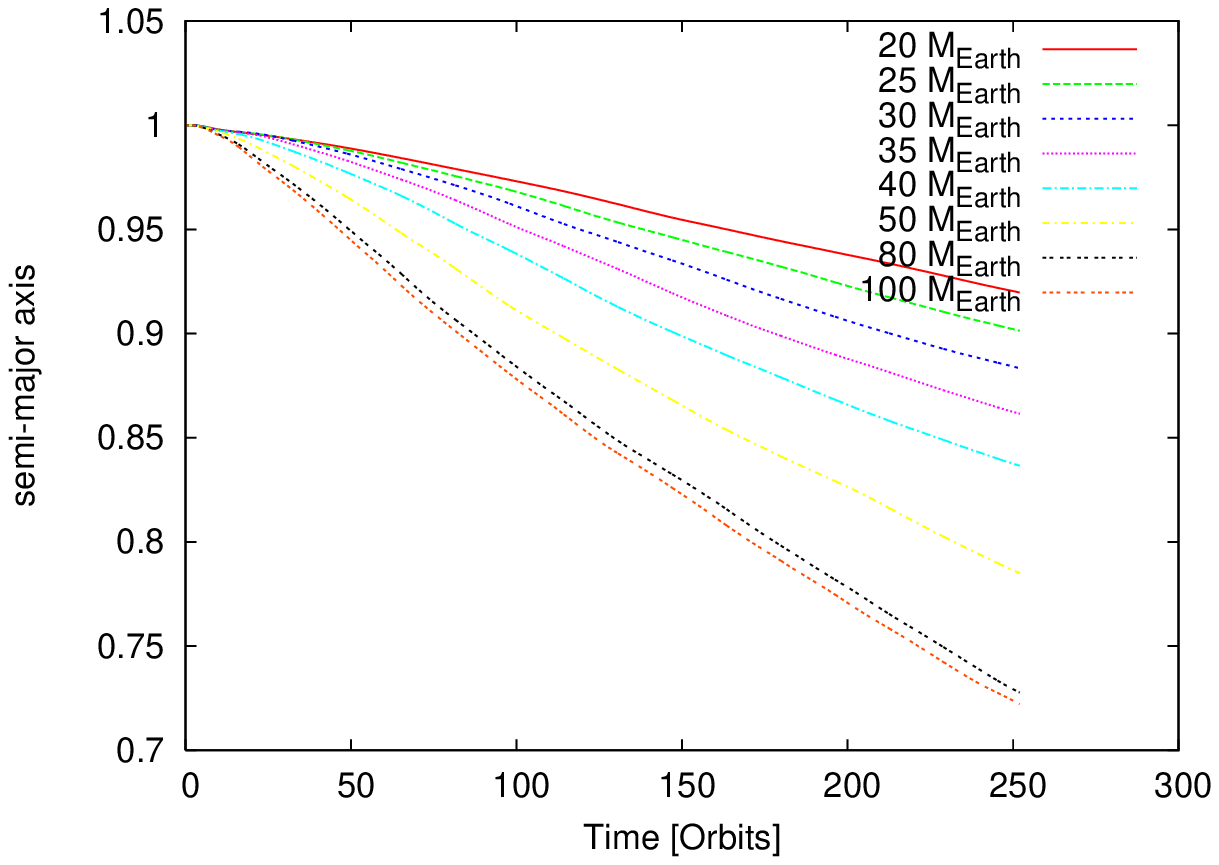}
 \includegraphics[width=0.9\linwx]{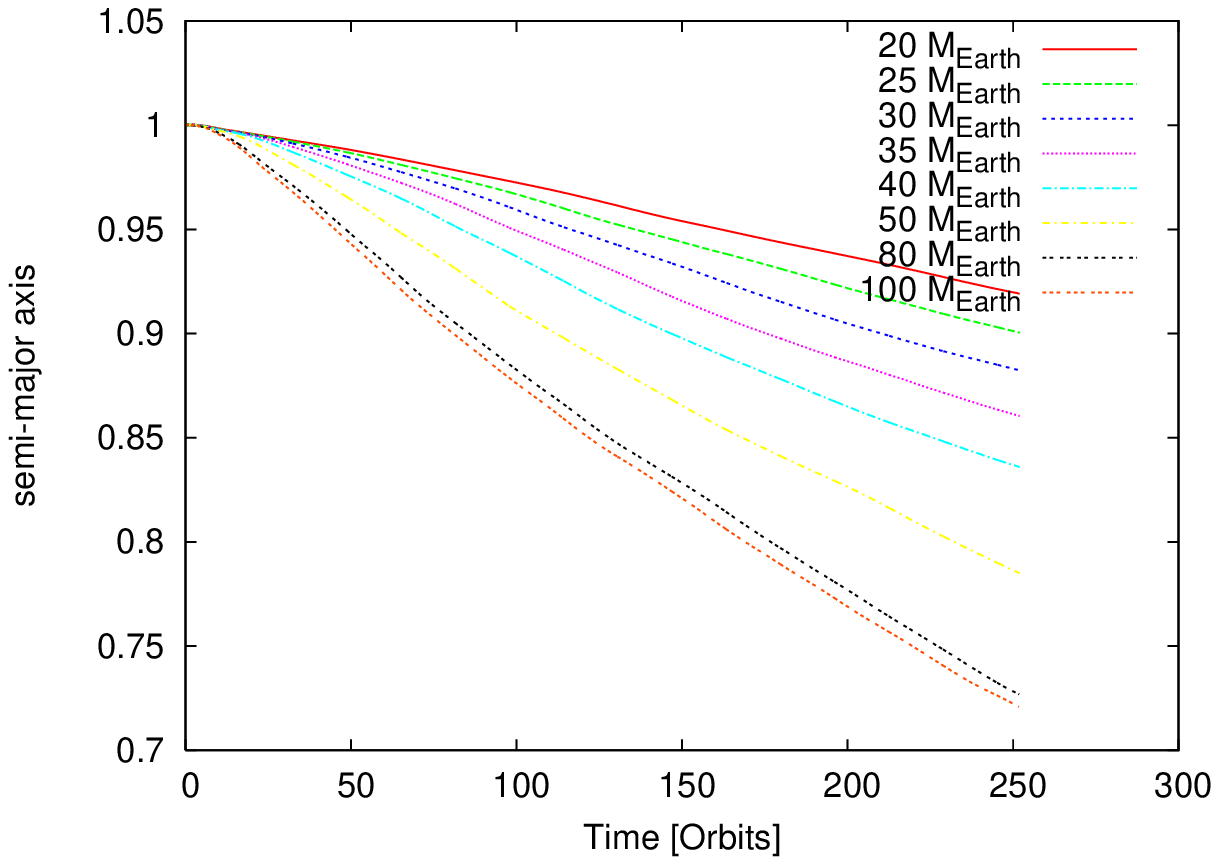}
 \caption{Time evolution of the semi-major axis for planets with $i_0 = 1.0^\circ$ (top) and $i_0 = 4.0^\circ$ (bottom) for different planetary masses in an isothermal $H/r=0.037$ disc.
   \label{fig:AInc1040Iso}
   }
\end{figure}

In Fig.~\ref{fig:AInc1040Iso} the evolution of the semi-major axis for the above mentioned planets is displayed. Interestingly it seems that the evolution of the semi-major axis does not depend on the inclination for circular orbits. All planets with equal masses follow the same evolution independent of the initial inclination. For the $20 M_{Earth}$ planet we found that the migration rate in the isothermal case does not depend much on the inclination of the planet, see Fig.~\ref{fig:AIncIsofullrate}. For higher mass planets this feature seems to be true as well. Of course, the planets migrate inwards at a speed dependent on the planetary mass. A higher planetary mass results in a more rapid inward migration of the planet, as it is predicted by linear theory \citep{2002ApJ...565.1257T}, although the theory formally applies only for an unperturbed disc. Our discs, however, show clear signs of perturbation in the surface density profile (Fig.~\ref{fig:Sig1040Iso}).

\begin{figure}
 \centering
 \includegraphics[width=0.9\linwx]{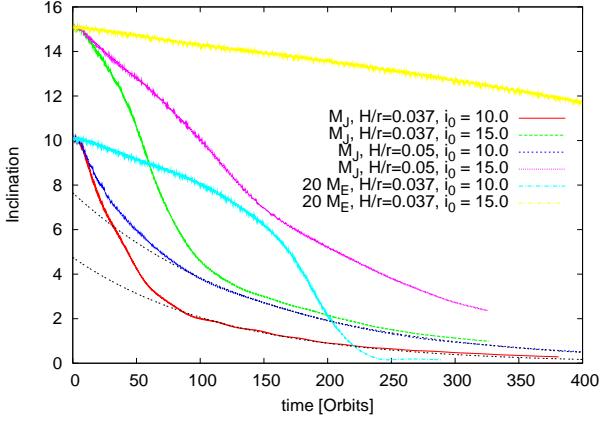}
 \caption{Time evolution of inclination for Jovian mass planets in isothermal $H/r=0.037$ and $H/r=0.05$ discs. The evolution of a $20 M_{Earth}$ planet in an isothermal $H/r=0.037$ disc is added for the inclination plot. The black dotted lines indicate a fit with $\tau_{inc}=120$ for a Jovian mass planet in an $H/r=0.037$ disc and a fit with $\tau_{inc}=145$ for a Jovian mass planet in a $H/r=0.05$ disc.
   \label{fig:IncJupIso}
   }
\end{figure}

In Fig.~\ref{fig:IncJupIso} the evolution the inclination for planets with a Jovian mass (and for an additional planet with $20 M_{Earth}$) in isothermal discs with different aspect ratios is displayed. The damping of inclination follows the predicted trend that the inclination of planets embedded in discs with smaller aspect ratio ($H/r=0.037$) is damped faster than the inclination of planets embedded in discs with higher aspect ratio ($H/r=0.05$), which was the result for smaller mass planets in our previous isothermal simulations as well. \citet{0004-637X-705-2-1575} stated that inclination damping of higher mass planets (e.g. Jovian mass) is considerably faster than for low mass planets (e.g. $20 M_{Earth}$). Fig.~\ref{fig:IncJupIso} 
shows clearly the same result for high inclinations ($i_0=15.0$), as the $20 M_{Earth}$ planet loses inclination at a much slower rate. 

However, for low inclinations ($i < 4.0^\circ$) the damping of inclination for Jovian mass planets is considerably slower than for low mass planets. The damping time scales obtained by fitting our simulations are considerably higher for Jovian mass planets in isothermal discs, in the $H/r=0.037$ disc $\tau_{inc, Jup}=120 > 23 = \tau_{inc, 20M_E}$ as well as in the $H/r=0.05$ disc $\tau_{inc, Jup}=145 > 53 = \tau_{inc, 20M_E}$. The $i_0=10.0^\circ$ planet with $20 M_{Earth}$ reflects this behaviour very well. First the damping of inclination is slower than for a Jovian mass planet, then the damping of inclination increases and the planet ends up in the midplane of the disc. The damping of inclination strongly depends on the interactions between disc and planet. 
As the Jovian mass planet opens a gap in the disc during the time of its evolution, the interactions between planet and disc are reduced, when the gap is opened. Therefore the damping of inclination of a Jovian mass planet is slowed down compared to a $20 M_{Earth}$ planet, when the inclination is already damped to about $i \approx 4.0^\circ$ after the planet has evolved for about $120$ orbits. After the gap has opened, only little mass remains adjacent to the planet to provide continuing damping. In the beginning of the simulations the gap has not opened yet and inclination damping is faster.

Overall, the damping of inclination starting from high inclinations ($i_0=15^\circ$) is faster for planets with a Jovian mass compared to small mass planets ($20 M_{Earth}$) as the damping time of inclination at high inclinations is much slower for low mass planets, which overcompensates the faster damping at low inclinations. This is in agreement with the statement of \citet{0004-637X-705-2-1575}. 

Inclination seems to have no visible effect on the migration of planets inside an isothermal disc, as long as the planet's orbit is circular. Eccentric orbits, on the other hand, tend to change the migration rate of the planet. Later on, we also investigate planets on inclined and eccentric orbits at the same time.

\subsection{Fully radiative disc}

\begin{figure}
 \centering
 \includegraphics[width=0.9\linwx]{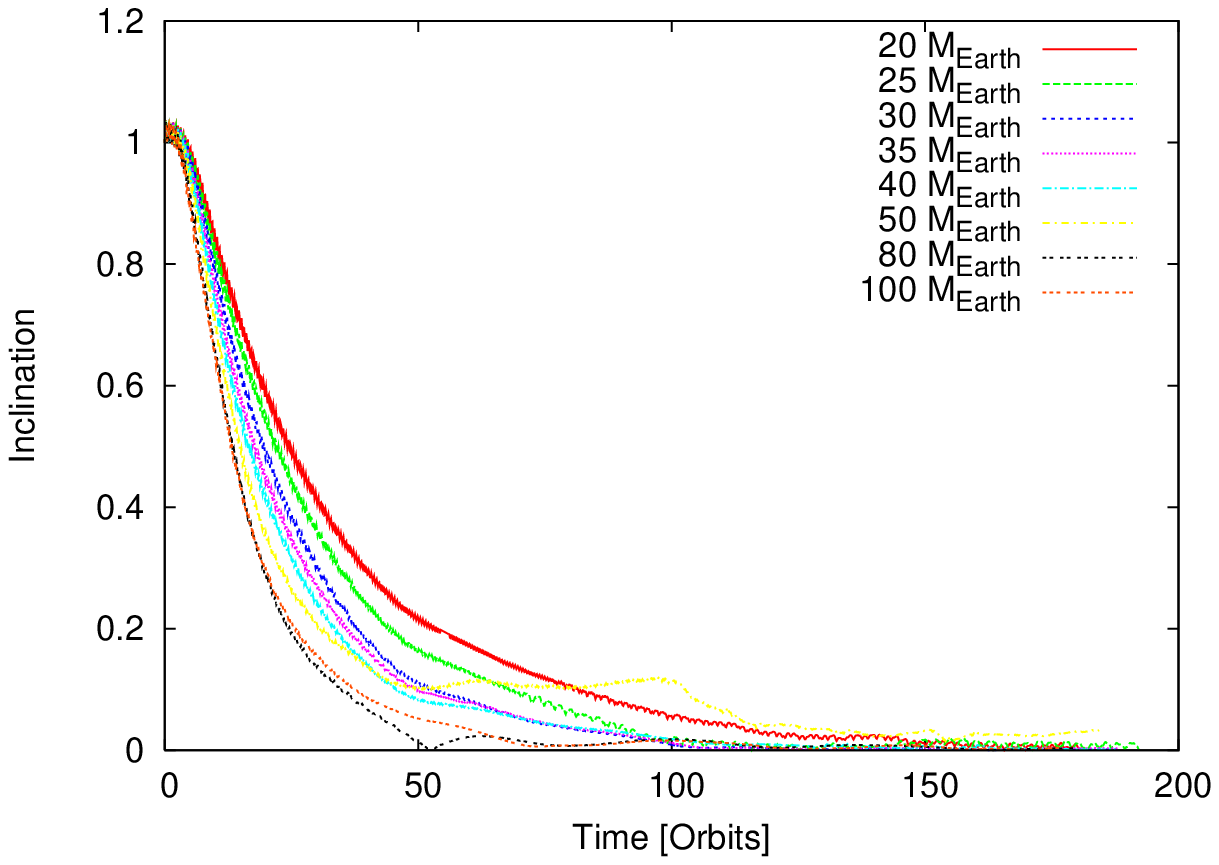}
 \includegraphics[width=0.9\linwx]{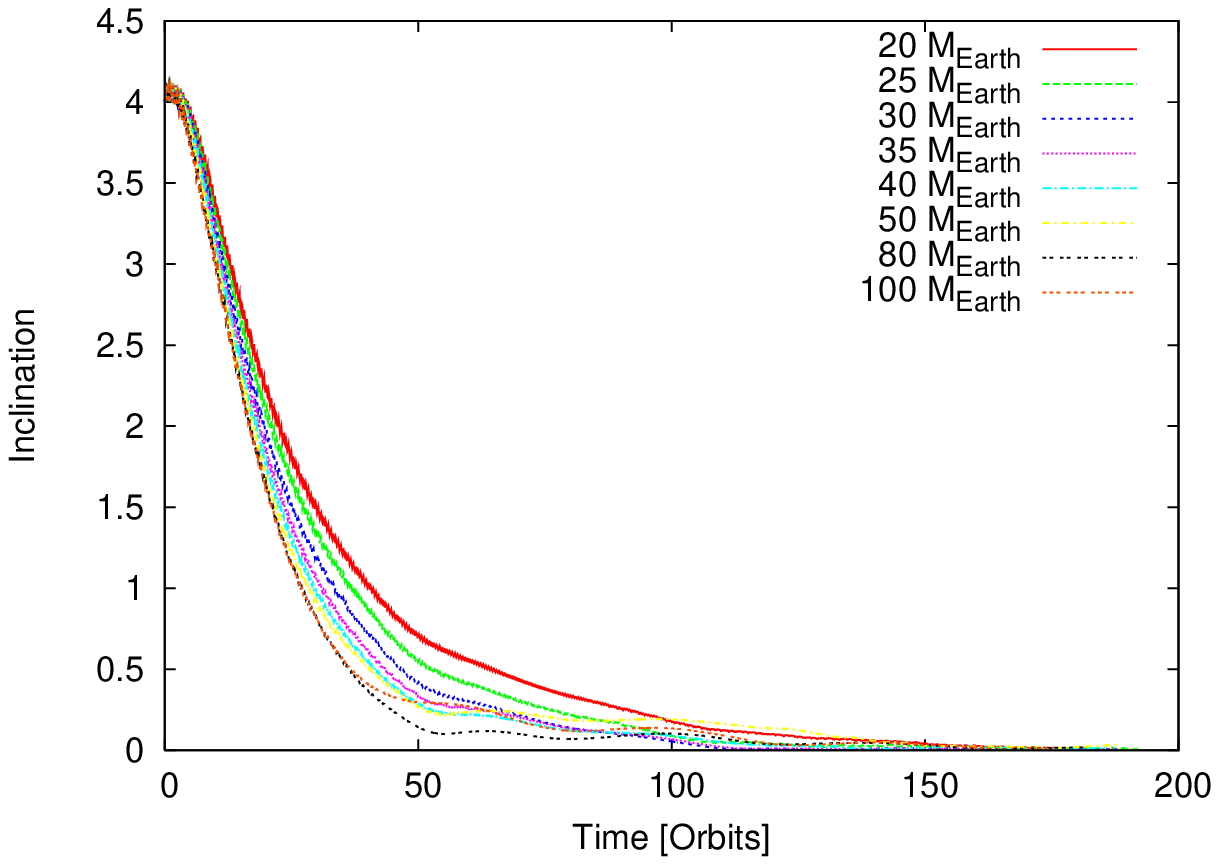}
 \caption{Time evolution of the inclination for planets with $i_0 = 1.0^\circ$ (top) and $i_0 = 4.0^\circ$ (bottom) for different planetary masses in a fully radiative disc.
   \label{fig:IncInc1040full}
   }
\end{figure}

In Fig.~\ref{fig:IncInc1040full} the evolution of inclination for planets with $i_0 = 1.0^\circ$ (top) and $i_0 = 4.0^\circ$ (bottom) for different planetary masses in a fully radiative disc is displayed. The inclination is damped for all planetary masses, but a higher planetary mass favours a faster damping of inclination for both starting inclinations. After about $150$ orbits the inclination of all planets with both individual starting inclinations is damped to approximately zero. The damping rate of the inclination is about the same for the fully radiative and the isothermal $H/r=0.037$ disc. 

In Fig.~\ref{fig:tauincIsofull} the measured damping time scale $\tau_{inc}$ is displayed for isothermal and fully radiative simulations. In the fully radiative case, the damping time scale is reduced for increasing planetary masses. If the planet reaches a certain mass (about $80 M_{Earth}$), the damping time scale is not reduced any more. As the planets start to open gaps in the disc, the damping of inclination is not further accelerated. It seems that gap opening planets do not only migrate inwards at the same speed, independent of the planets mass, but also damp their inclination at the same speed.

\begin{figure}
 \centering
 \includegraphics[width=0.9\linwx]{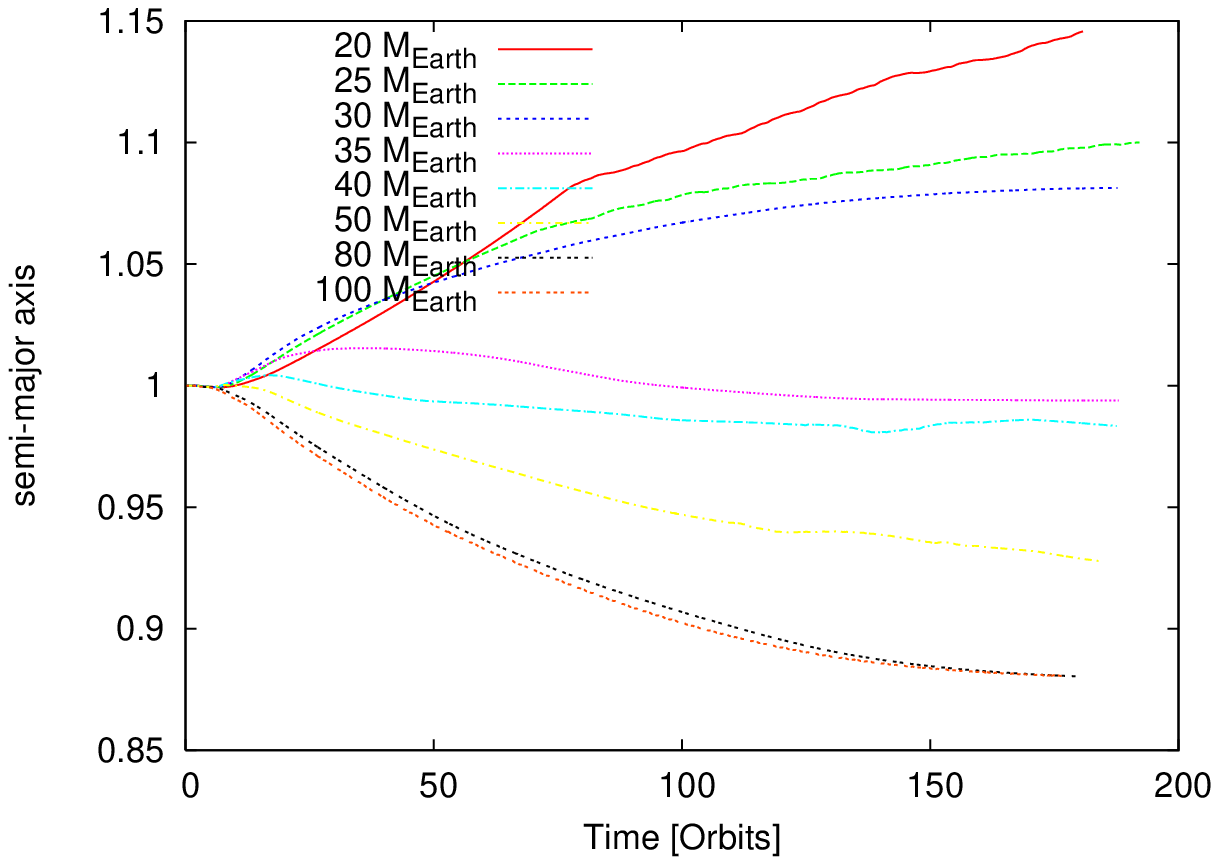}
 \includegraphics[width=0.9\linwx]{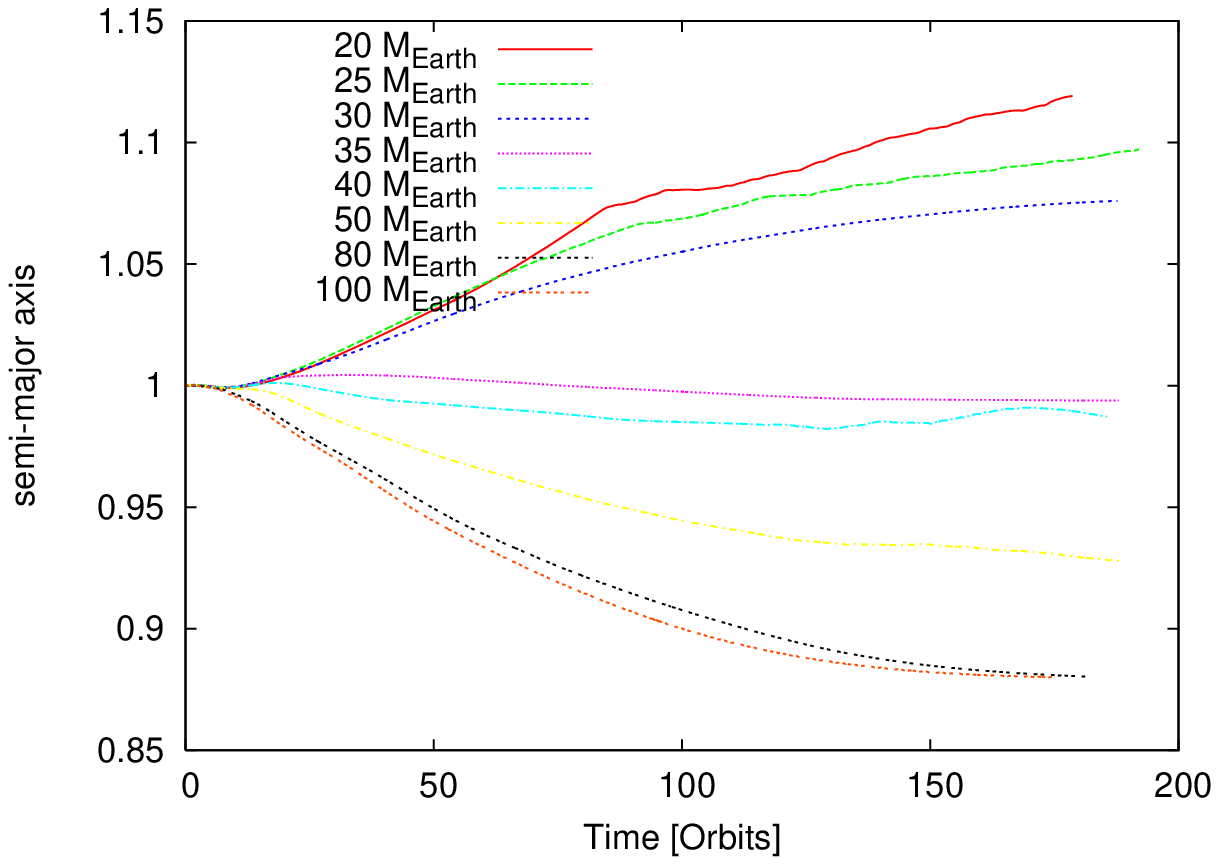}
 \caption{Time evolution of the semi-major axis for planets with $i_0 = 1.0^\circ$ (top) and $i_0 = 4.0^\circ$ (bottom) for different planetary masses in a fully radiative disc.
   \label{fig:AInc1040full}
   }
\end{figure}

In Fig.~\ref{fig:AInc1040full} the evolution of the semi-major axis for planets with $i_0 = 1.0^\circ$ (top) and $i_0 = 4.0^\circ$ (bottom) for different planetary masses in a fully radiative disc is displayed. As expected from our previous simulations \citep{2009A&A...506..971K} only planets with a planetary mass of $M_P \leq 33 M_{Earth}$ migrate outwards, while heavier objects migrate inwards. The $35$ and $40 M_{Earth}$ planets seem to migrate inwards just a little bit, before their migration is nearly stopped. The higher mass planets ($50$, $80$ and $100 M_{Earth}$) migrate inwards at nearly the same speed as the planets inside the isothermal $H/r=0.037$ disc.
These observations are independent of the planets inclination.

The outward migrating low mass planets travel outwards at a slightly faster speed in the $i_0=1.0^\circ$ case compared to the $i=4.0^\circ$ simulations. This can be expected, as the calculated migration rate for $20 M_{Earth}$ planets on fixed orbits is about a factor of $5$ to $8$ higher in the low inclined case. The higher migration rate for the lower inclined planets results in a faster outward migration. Finding the observed difference in the semi-major axis being so small after about $150$ orbits is a result of inclination damping. As the inclination is damped, the migration rate increases significantly until the inclination is $\approx 2.0^\circ$, which takes only a few orbits starting from $i_0=4.0^\circ$. In this little time frame, the planet is not able to migrate a large distance, so we only observe a small difference in the evolution of the semi major axis.

\section{Inclined and eccentric planets}
\label{sec:inceccplanets}

Inclination did not change the direction of migration for planets on circular orbits. Eccentricity on the other hand stopped the outward migration of low mass planets in fully radiative discs, as long as the eccentricity of the planet is higher than $e=0.02$ \citep{2010A&A.523...A30}. However, planets with lower eccentricity migrate outwards in the fully radiative scheme. If an inclined planet surrounds a star on an eccentric orbit, in what way will these two properties of the orbit influence each other? To find an answer, simulations of a low mass planet ($20 M_{Earth}$) on an inclined and eccentric orbit in an isothermal $H/r=0.037$ and in a fully radiative disc are computed.

\subsection{Isothermal disc}

Our recent work with planets on eccentric orbits has shown that for isothermal simulations with moving planets on eccentric orbits the cubic potential (for the planetary potential) should be replaced by the $\epsilon$-potential. In the following, isothermal simulations of planets on eccentric and inclined orbits we use the $\epsilon$-potential with a smoothing length of $r_{sm}=0.8$. 

\begin{figure}
 \centering
 \includegraphics[width=0.9\linwx]{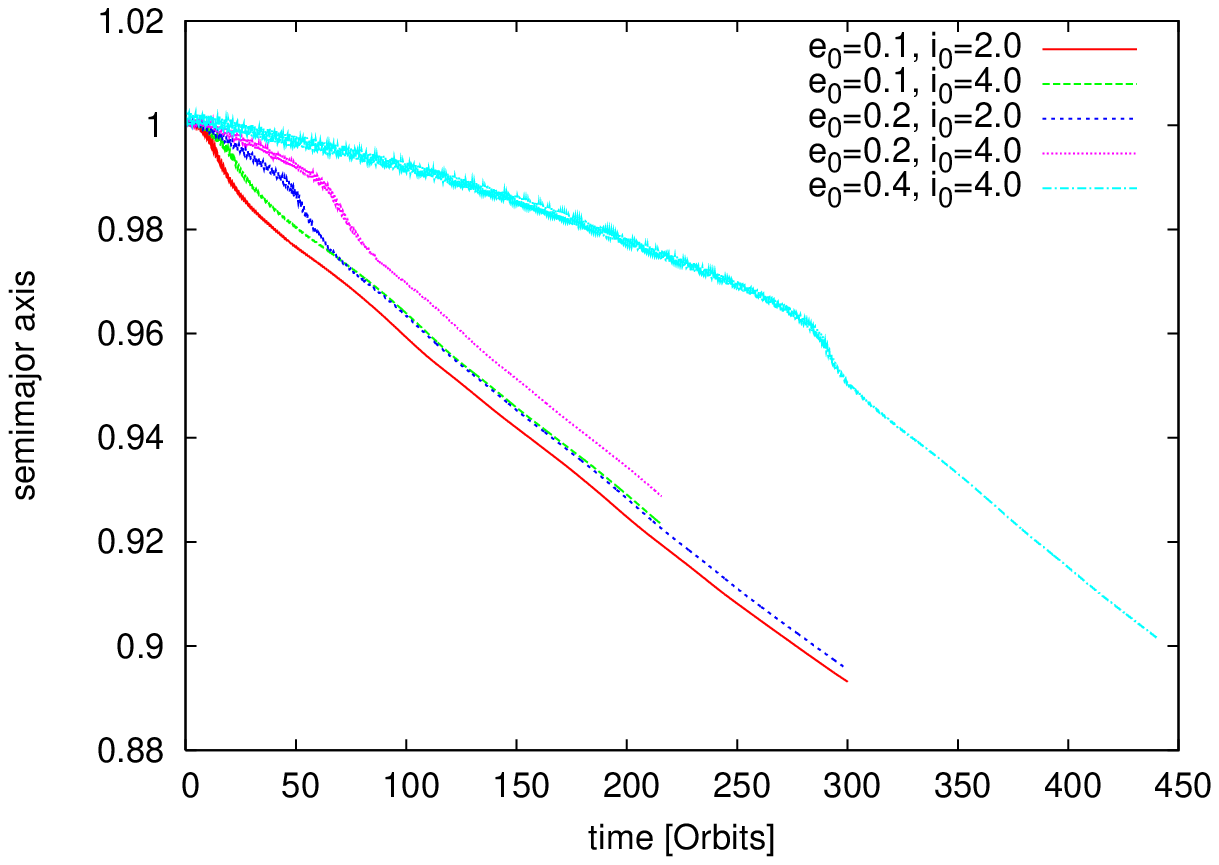}
 \includegraphics[width=0.9\linwx]{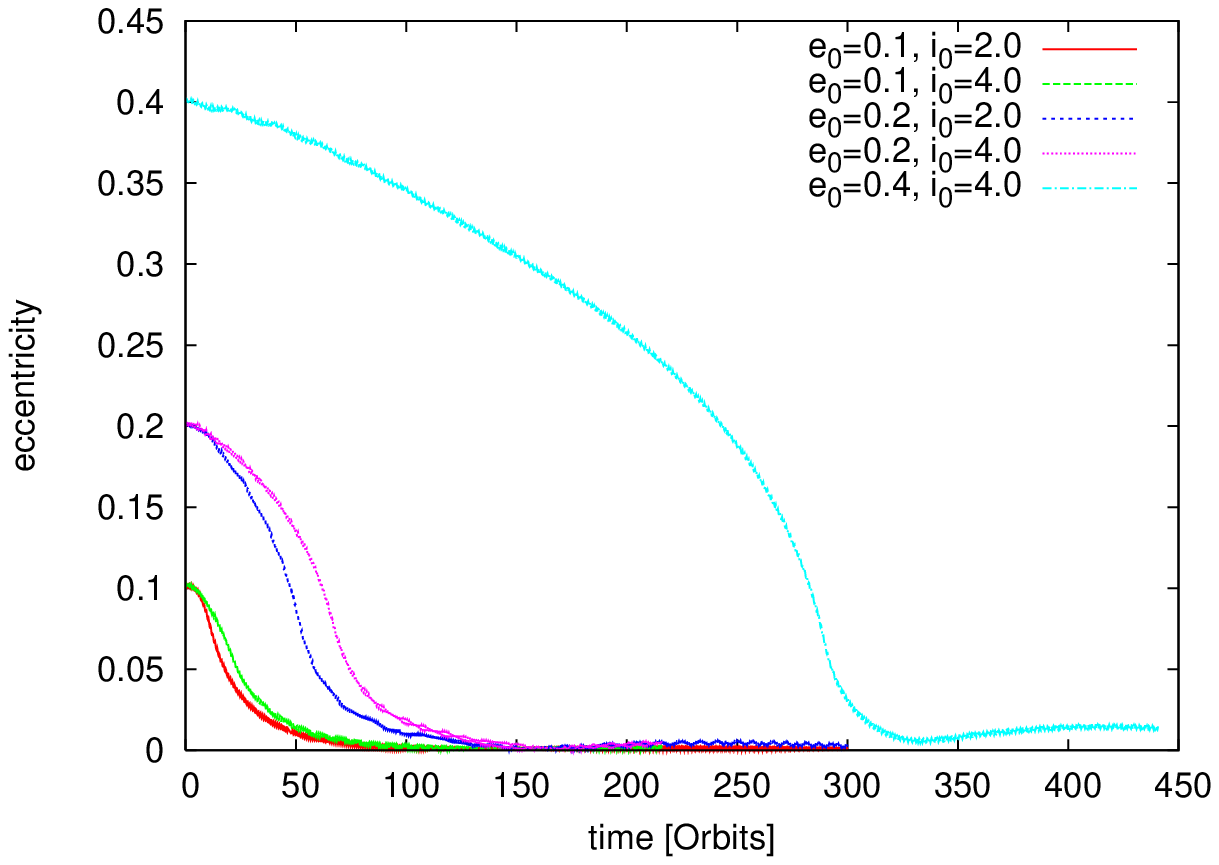}
 \includegraphics[width=0.9\linwx]{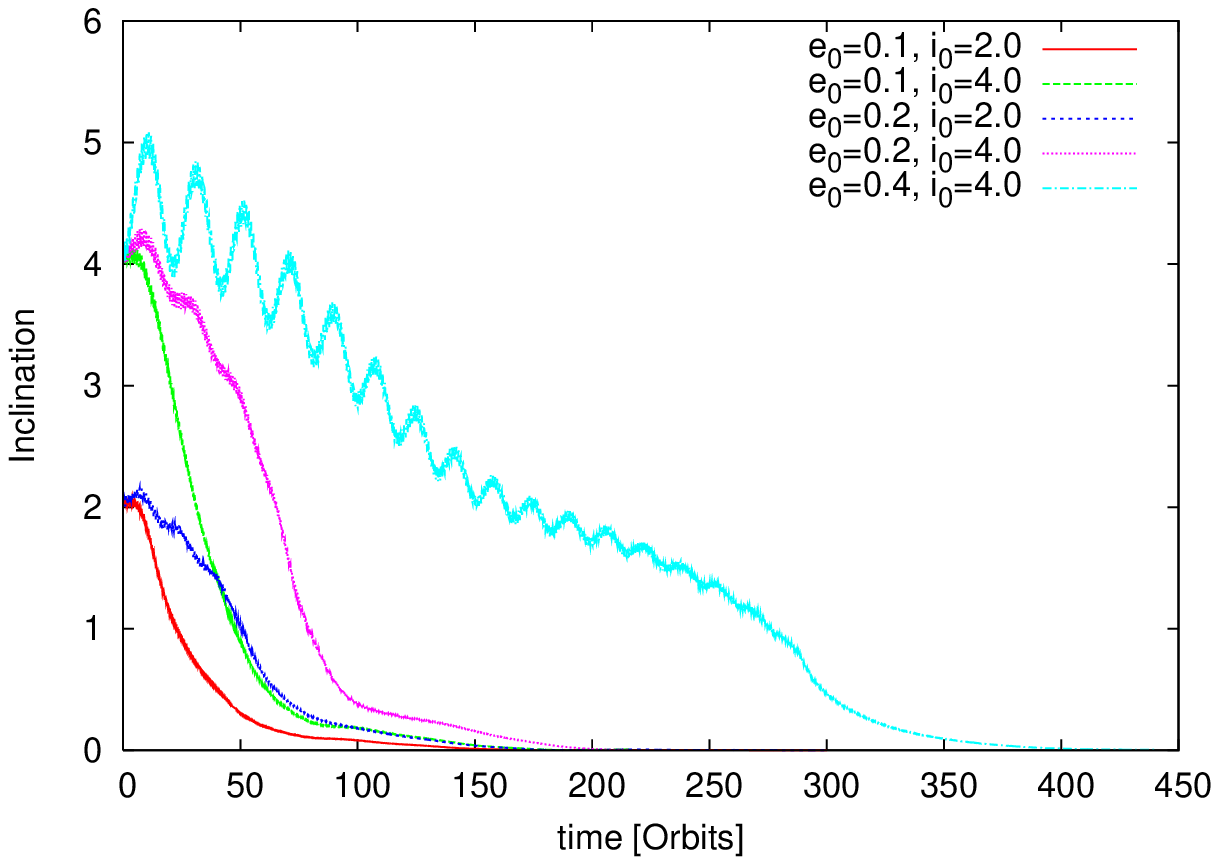}
 \caption{Time evolution of semi-major axis (top), eccentricity (middle) and inclination (bottom) for planets ($20 M_{Earth}$) with different eccentricity and inclination in an isothermal $H/r=0.037$ disc.
   \label{fig:AExcIncIso}
   }
\end{figure}

In this work we limit ourselves to planets with an initial eccentricity of $e_0=0.1$, $e_0=0.2$ and $e_0=0.4$ and with an initial inclination of $i_0=2.0^\circ$ and $i_0=4.0^\circ$. In Fig.~\ref{fig:AExcIncIso} the evolution of the semi-major axis (top), the eccentricity (middle) and inclination (bottom) for planets in an isothermal $H/r=0.037$ disc is displayed. 

The planets migrate inwards (loss of semi-major axis) as expected for low-mass planets in an isothermal disc. For planets with an initial inclination of $i_0=4.0^\circ$, the planets migrate inwards at a slower rate compared to the lower inclined cases. A lower initial eccentricity results in a faster inward migration in the beginning of the simulation for all inclinations. This faster inward migration seems to be in a direct correlation with the planet's eccentricity \citep{2010A&A.523...A30}. When the planets eccentricity reaches about $e \approx 0.1$, the planet undergoes a short rapid inward migration and a more rapid loss of eccentricity, compared to the loss of eccentricity at different eccentricities. As soon as the $e_0=0.2$ planet's eccentricity is damped to about $e \approx 0.1$, the planet undergoes a rapid inward migration leading to nearly the same evolution of semi-major axis as the $e_0 = 0.1$ planet from this time on, again for both inclinations. Interestingly, a lower starting inclination results in a faster inward migration, when the planets have an initial eccentricity, compared to planets on circular orbits, where the inclination seems to have no effect on the migration. A higher initial inclination ($i=4.0^\circ$) seems to result in the opposite: a slower inward migration of the planet. We have observed a similar effect for eccentricities: planets with a lower initial eccentricity seem to migrate inwards at a slower rate compared to their highly eccentric counterparts \citep{2010A&A.523...A30}.

The initial eccentricity is damped for all these simulations with only two small differences. Planets with an initially higher eccentricity need a longer time to be damped to zero and planets with the same eccentricity but a higher inclination also need a longer time to damp their eccentricity. As the eccentricity is damped to lower values, at $e \approx 0.1$ we observe a more rapid loss of eccentricity compared to the loss of eccentricity at other times. At this point the planet loses about $50\%$ of its (actual) eccentricity in the time of a few orbits. At this point in time we also observe a faster inward migration of the planet. This dependency of a rapid drop of eccentricity and a resulting decrease of semi-major axis in the isothermal regime is in agreement with the results found in \citet{2007A&A...473..329C} and \citet{2010A&A.523...A30}

The evolution of the inclination for the low eccentric planet ($e_0=0.1$) follows nearly the evolution of a planet on circular orbit. The damping time is in fact a marginally longer for the eccentric case, but besides this effect there is no difference. For the $e_0=0.2$ case the inclination starts to oscillate slightly in the beginning of the simulation until it is damped to about $i \approx 3.0^\circ$ or $i \approx 1.5^\circ$. Then the inclination follows approximately the damping rate of a planet on a circular orbit (see Fig.~\ref{fig:IncIncIso}) until it is damped to about zero. The high eccentric case ($e_0=0.4$) shows even more and stronger oscillations in the inclination. In the beginning the planets inclination is pumped up to about $\approx 5.0^\circ$, which is $25\%$ more than the starting inclination. In time, the amplitude of the oscillations becomes smaller and the frequency higher while the overall inclination dissipates in time. These oscillations are observed 
 in \citet{2007A&A...473..329C} as well. The time until the inclination reaches zero is about $4$ times as large as in the zero eccentricity case.

By comparing the migration rate of the inclined and eccentric planet with only an eccentric planet (migrating in midplane with a zero inclination), we observe nearly the same migration rate for both planets. Due to these results, it seems that inclination does not influence the migration in an isothermal disc at all.

\subsection{Fully radiative disc}

\begin{figure}
 \centering
 \includegraphics[width=0.9\linwx]{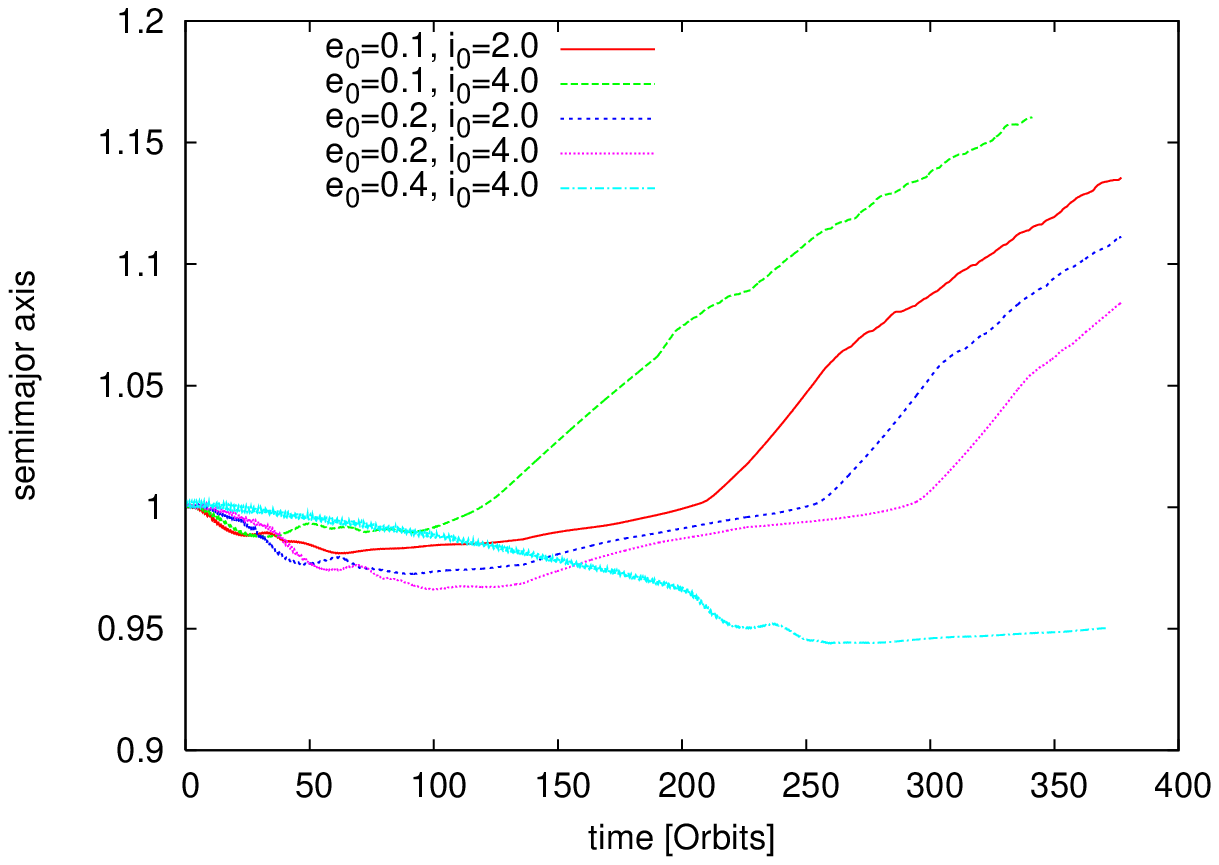}
 \includegraphics[width=0.9\linwx]{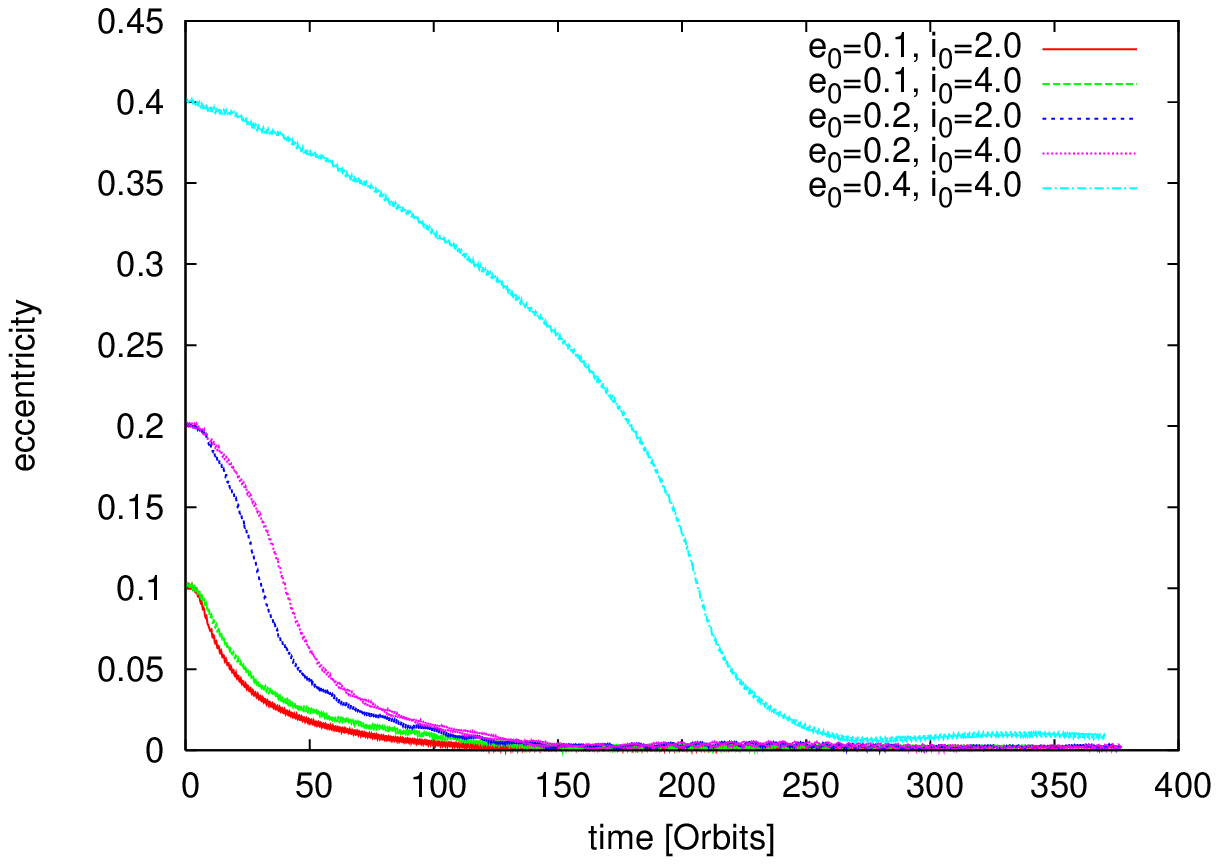}
 \includegraphics[width=0.9\linwx]{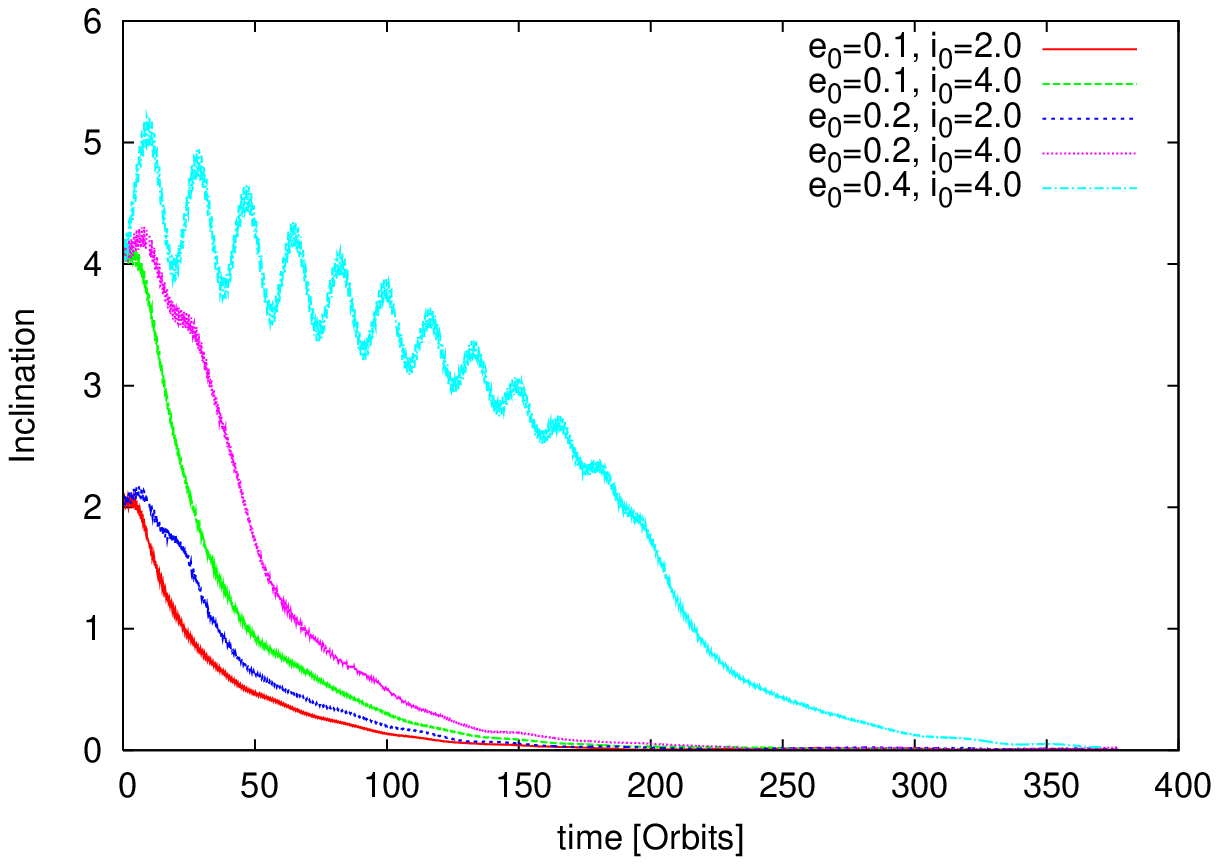}
 \caption{Time evolution of semi-major axis (top), eccentricity (middle) and inclination (bottom) for planets ($20 M_{Earth}$) with different eccentricity and inclination in a fully radiative disc.
   \label{fig:AExcIncfull}
   }
\end{figure}

For the fully radiative simulations we rely on our cubic $r_{sm}=0.5$ potential again. For comparison, the starting eccentricity and inclination for the fully radiative simulations match those of the isothermal simulations. In Fig.~\ref{fig:AExcIncfull} the evolution of the semi-major axis (top), the eccentricity (middle) and inclination (bottom) for planets in a fully radiative disc are displayed. 

In the beginning of the simulations all planets migrate inwards in the same way as in the isothermal simulations, but in contrast to the isothermal simulations the planets reverse their inward migration and migrate outwards after about $\approx 100$ orbits in the low eccentric case ($e_0 \leq 0.2$). The high eccentric case ($e_0=0.4$) takes a much longer time to reverse its inward migration. A higher initial eccentricity results in a later outward migration of the planet. The evolution of the semi-major axis contains at some point a quite fast loss of semi-major axis for the initial low eccentric planets. This loss of semi-major axis correlates to the eccentricity. As soon as the eccentricity is damped to $\approx 0.1$, the fast loss of the semi-major axis sets in. The inclination of the planet seems to play no significant role in the migration process of these simulations, as for the $e_0=0.1$ case the higher inclined planet ($i_0=4.0^\circ$) migrates outward first, while for the $e_0=0.2$ case the lower inclined planet ($i_0=2.0^\circ$) migrates outward first.

The initial eccentricity is damped for all planets, but for planets with the same starting eccentricity, the time needed to damp the eccentricity is longer for the higher inclined planets, see the middle of Fig.~\ref{fig:AExcIncfull}. As the eccentricity is damped to $\approx 0.1$, it is damped with an exponential decay, which seems to result in a faster loss of semi-major axis at the same time in the evolution of the planet. The damping for the initial low eccentric planets ($e \leq 0.2$) takes about $120$ planetary orbits, while a higher initial eccentricity ($e_0=0.4$) takes about three times longer.

The inclination damping follows in principle the evolution of inclination of the isothermal simulations. As soon as the planet is released in the disc, the inclination is damped as long as the eccentricity is very low ($e_0=0.1$). Higher eccentricities seem to affect the inclination in such a way that the inclination begins to oscillate. This oscillation is already visible in the $e_0=0.2$ case, but becomes much stronger for even higher eccentricities. The oscillations begin to dissipate as soon as the planets eccentricity is damped to about $e \leq 0.1$. During time these oscillations lose amplitude and gain frequency (as we can see from the $e_0=0.4$ simulation). A higher eccentricity seems to cause a higher and more frequent oscillation of the inclination. These oscillations have been observed in \citet{2007A&A...473..329C} as well.

\section{Summary and Conclusions}
\label{sec:Sumcon}

We performed full 3D radiation hydrodynamical simulations of accretion discs with embedded planets of different masses on inclined orbits. In a first sequence of simulations we have analysed in detail the dynamics of a planet with a given mass of $20 M_{Earth}$ for the isothermal as well as the fully radiative case. For planets on fixed orbits we calculate the expected change of inclination and semi-major axis of an inclined planet on a circular orbit embedded in the disc, which we confirm subsequently through simulations of moving planets. For the isothermal ($H/r=0.037$ and $H/r=0.05$) as well as for the fully radiative simulations the inclination is damped in time. The damping time scale in isothermal discs depends on the disc's thickness ($H/r$), as the inclination is damped faster for discs with lower aspect ratio than in discs with higher aspect ratio. A smaller $H/r$ also results in a faster inward migration compared to a higher $H/r$. Our simulation results agree in this respect with \citet{2002ApJ...565.1257T}. 

For planets on inclined orbits we confirm the outward migration. Our results for planets on fixed orbits agree very well with the simulations of planets moving inside the disc. We find that outward migration interestingly occurs over a relatively large range of inclination, up to about ($i=4.5^\circ$), with a faster outward migration for lower inclined planets. For higher inclined planets the migration seems to be stalled in general.
In our simulations, the damping of inclination is slower in the fully radiative disc ($\tau_{inc}=26.5$) compared to the isothermal disc having the same scale height ($\tau_{inc}=23$), a difference that may be attributed to the different sound speeds, isothermal versus adiabatic.

As recent studies showed, planets in a fully radiative disc only migrate outwards if they do not exceed a certain critical mass \citep{2009A&A...506..971K},  beyond which gap formation sets it.
This outward migration does crucially depended on the shape of the orbit of the planet in the disc, as discussed in \citet{2010A&A.523...A30, 2009A&A...506..971K}. Here, we extended those studies to inclined planets and found that planets on circular and inclined orbits migrate inwards in the isothermal case, and migrate outwards (if the planetary mass is low enough and the inclination of the planet is below the threshold of $i\approx 4.5^\circ$) in the fully radiative case. For higher mass planets, our previous results have been confirmed in the expected way. Planets with a mass up to $\approx 33 M_{Earth}$ migrate outwards and planets with higher masses migrate inwards in the fully radiative scheme. In the isothermal case, planets migrate inwards as predicted by linear theory \citep{2002ApJ...565.1257T}. 
When evolving planets with higher planetary masses, the simulations started from unperturbed discs, which
changes the damping rate of inclination compared to planets starting in perturbed discs (by the presence of the embedded planet).
This was also stated for inclined high mass planets in isothermal discs by \citet{0004-637X-705-2-1575}. After the inclination is damped in time, the direction of migration only depends on the thermodynamics of the disc and on the planetary mass. 

For eccentric and inclined low mass planets we observe a quite different behaviour. Eccentricity seems to slow down the inward migration of planets in the isothermal regime and prevents outward migration in the fully radiative scheme (for low mass planets which would normally migrate outward). Since eccentricity is damped with time, this leads to slightly faster inward migration in the isothermal regime and to a reverse of the migration direction in the fully radiative scheme. The origin of this outward migration is created by the delicate density structure near the planet, which is destroyed by even a very low amount of eccentricity \citep{2010A&A.523...A30}. A high initial eccentricity gives rise to oscillations in the inclination, which become weaker and more frequent in time, as the eccentricity is damped. These oscillations seem to be responsible for a slower damping of inclination compared to low eccentric planets. 

The results by \citet{0004-637X-705-2-1575} for Jovian mass planets indicated that inclination and eccentricity is damped in a very short time compared to low mass planets. We can confirm their results for inclination damping in discs with two different aspect ratios, however this is only true for the damping of high inclinations. For low inclinations ($i \leq 10^\circ$) the damping of inclination for low mass planets is faster than for high mass planets. However, when considering the total time needed to damp the inclination of a planet starting from $i_0=15.0^\circ$, the damping is much faster for high mass planets, which confirms the results by \citet{0004-637X-705-2-1575}. A Jovian mass planet with high inclination evolves more rapidly as its interactions with the disc is stronger if no gap has opened yet.

Our results indicate that the inclination as well as the eccentricity of single planets are damped by the disc.
At the end of our simulations the planets end up in midplane, meaning they have no remaining inclination.
While this is in agreement with the very flat solar system, it seems to be in contradiction to the observed
highly inclined exoplanets \citep{2010arXiv1008.2353T}.
However, it is known that higher eccentricities as well as inclinations can be produced via planet-planet
interaction.
For a pair of planets embedded in discs convergent migration processes can excite
eccentricity to high values depending on the damping action of the disc \citep{2008A&A...483..325C}.
Stronger scattering events can be induced in case of a multiple planet system still embedded in the
disc \citep{2010A&A...514L...4M}. During the subsequent longterm evolution these initially excited system
may then evolve into configurations with very high eccentricity as well as inclination 
\citep{2008ApJ...686..580C,2010ApJ...714..194M}. Thus, the combination of initial planet-disc interaction
with subsequent scattering processes and tidal interaction with the central star is one pathway
towards the observed
strongly tilted systems \citep{2010ApJ...718L.145W}.
On the other hand, systems such as Kepler-9 \citep{2010Sci...330...51H} seem to show clear evidence
for continued migration towards the star.  

\begin{acknowledgements}

B. Bitsch has been sponsored through the German D-grid initiative. W. Kley acknowledges the support through the German Research Foundation (DFG) through grant KL 650/11 within the Collaborative Research Group FOR 759: {\it The formation of Planets: The Critical First Growth Phase}. The calculations were performed on systems of the Computer centre of the University of T\"ubingen (ZDV) and systems operated by the ZDV on behalf of bwGRiD, the grid of the Baden  W\"urttemberg state. Finally, we gratefully acknowledge the helpful and constructive comments of an anonymous referee.

\end{acknowledgements}

\bibliographystyle{aa}
\bibliography{kley8}
\end{document}